\documentclass[preprint]{aastex}

\usepackage{graphicx}
\usepackage{subfigure}
\usepackage{latexsym}

\newcommand{\HI}{\protect\ion{H}{1}}

\newcommand{\msun}{$M_\odot$}
\newcommand{\lsun}{$L_\odot$}
\newcommand{\etal}{{et~al.}}
\newcommand{\cmsq}{cm$^{-2}$}
\newcommand{\mhi}{$M_{HI}$}
\newcommand{\nhi}{$N_{HI}$}
\newcommand{\kms}{~km\,s$^{-1}$}

\newcommand{\vgsr}{$v_{\rm GSR}$}

\begin{document}

\date{\today}

\title{An \HI\ Survey of Six Local Group Analogs.  II.  \HI\ properties of group galaxies}
\author{D.J. Pisano\altaffilmark{1}}
\affil{Department of Physics, West Virginia University, P.O. Box 6315, Morgantown, WV, 26506, 
USA}
\email{djpisano@mail.wvu.edu}
\altaffiltext{1}{Adjunct Assistant Astronomer at National Radio Astronomy Observatory, 
P.O. Box 2, Green Bank, WV 24944}
\author{David G. Barnes}
\affil{Centre for Astrophysics \& Supercomputing, Swinburne University, 
Hawthorn, Victoria 3122, Australia}
\email{David.G.Barnes@gmail.com}
\author{Lister Staveley-Smith}
\affil{International Centre for Radio Astronomy Research, M468, University of Western Australia, Crawley, WA 6009, Australia}
\email{Lister.Staveley-Smith@icrar.org}
\author{Brad K. Gibson}
\affil{Jeremiah Horrocks Institute, University of Central Lancashire, Preston, 
PR1 2HE, UK}
\affil{Department of Astronomy \& Physics, Saint Mary's University, Halifax,
Nova Scotia, B3H 3C3, Canada}
\email{brad.k.gibson@gmail.com}
\author{Virginia A. Kilborn}
\affil{Centre for Astrophysics \& Supercomputing, Swinburne University, 
Hawthorn, Victoria 3122, Australia}
\email{vkilborn@astro.swin.edu.au}
\author{Ken C. Freeman}
\affil{RSAA, Mount Stromlo Observatory, Cotter Road, Weston, ACT 2611, 
Australia}
\email{kcf@mso.anu.edu.au}

\begin{abstract}

We have conducted an \HI\ 21 cm emission-line survey of
six loose groups of galaxies chosen to be analogs to the Local Group.  The
survey was conducted using the Parkes Multibeam instrument and the Australia 
Telescope Compact Array (ATCA) over a $\sim$1 Mpc$^2$ area and covering the 
full depth of each group, with a \mhi\ sensitivity of $\sim$7$\times$10$^5$\msun.  
Our survey detected 110 sources, 61 of which are associated with the six groups.  All of these
sources were confirmed with ATCA observations or were previously cataloged by
HIPASS.  The sources all have optical counterparts and properties consistent with
dwarf irregular or late-type spiral galaxies.  We present here the \HI\ properties of the 
groups and their galaxies.  We derive an \HI\ mass function for the groups that is consistent 
with being flatter than the equivalent field HIMF.  We also derive a circular velocity distribution function, tracing the luminous dark 
matter halos in the groups, that is consistent with those of the Local Group and HIPASS 
galaxies, both of which are shallower than that of clusters or predictions from CDM models of galaxy formation.
\end{abstract}


\section{Introduction}

The majority of galaxies, including the Milky Way, reside in groups
\citep{geller83,tul87,eke04,tago08}, as such it is essential to study these structures if we wish
to understand the effect of the environment on galaxy properties.  A galaxy group is a
very broad classification that has not been very precisely defined in the literature and whose
properties span a wide range of mass and density (amongst others).  They range in size
from massive, rich groups to low mass poor, loose groups and compact groups.  The rich groups
tend to be dominated by early-type galaxies \citep{postman84, helsdon03} and have an X-ray bright, 
intra-group medium \citep[IGM;][]{mulchaey00,mulchaey03} that may result in ram pressure 
stripping of gas-rich spiral galaxies \citep{sengupta06,sengupta07}.  In these ways, rich groups are 
very similar to low-mass galaxy clusters.  Compact groups are the densest groups with a range of 
masses, containing a few to tens of galaxies typically separated by only a couple of galaxy radii 
\citep[e.g.][]{hickson82,hickson97}.  These groups contain galaxies that are strongly interacting
and can also host an X-ray bright IGM  \citep{ponman96} potentially generated by the tidal 
interactions of the group members \citep{v-m01}.  

In contrast to both of these classes, loose, poor groups are similar to the Local Group.  
They are less massive than rich groups, although with similar numbers of member galaxies.  They
can be dominated by either late-type or early-type galaxies, but only the groups containing at least
one early-type galaxies have a hot, X-ray emitting IGM \citep{mulchaey00,mulchaey03}.  
This suggests that ram pressure stripping is unlikely to have a large effect on galaxies in most of
these groups \citep[cf.][]{grcevich09}.  These groups are diffuse with low velocity 
dispersions resulting in crossing times that are comparable to a Hubble time and, as such, they are unlikely to 
be virialized \citep{zab98}.  This implies that interactions are rare, but, due to the low velocity 
dispersions, are more effective than in clusters and rich groups.  These interactions may even strip 
\HI\ from galaxies in loose, poor groups \citep{omar05}.

While there have been a number of studies of the gaseous properties of isolated galaxies 
\citep{haynes84,pis99,pis02}, of cluster galaxies \citep[e.g.][]{bravo-alfaro00,
bravo-alfaro01,chung09}, and of compact and rich groups \citep[e.g.][]{williams87, williams91, 
v-m01, freeland09, kilborn09, borthakur10}, there have been fewer targeted studies of poor, loose 
groups analogous to our own Local Group.  This paper seeks to fill that gap by exploring the 
neutral hydrogen properties of galaxies in six nearby, loose groups.  

Mass functions serve as an excellent test of models of galaxy formation and are a simple of way
quantifying differences between galaxy populations in different environments and comparing 
observations with models \citep{snaith11}.  Current models of 
cold dark matter galaxy formation predict an excess of low mass, as parameterized by their circular velocity 
(the circular velocity distribution function; CVDF), dark matter halos as compared to what 
is observed locally \citep{klypin99,moore99}.  While some authors have directly measured the 
CVDF for luminous galaxies \citep[e.g.][]{shimasaku93, sheth03,goldberg05}, this is difficult 
to do for very low mass dwarf galaxies; \citet{blanton08} and others have used different methods to
infer the CVDF or used alternative proxies.  The optical luminosity function has been regularly used 
as a proxy \citep[e.g.][]{tully02,trentham05} for the CVDF.  Luminosity functions can also be used to
in concert with the Tully-Fisher relation or Fundamental Plane to infer the CVDF 
 \citep{cole89, gonzalez00,sheth03,desai04,goldberg05}.  Unlike optical luminosity functions using
radio observations of 21-cm emission from neutral hydrogen (\HI) provides two measures of the 
mass of a galaxy:  the integrated line profile yields the \HI\ mass while the linewidth provides the
circular rotation velocity of the galaxy.  Regardless of how the halo mass function is measured, be
it by optical luminosity \citep{tully02, trentham05}, \HI\ mass \citep{zwaan05}, or circular velocity 
\citep{zwaan10}, there is always a deficit of observed galaxies at low masses; this is the ``missing satellite" 
or ``substructure" problem \citep{klypin99,moore99}.  Furthermore, there have been relatively
few studies of how the \HI\ mass function (HIMF) and circular velocity distribution 
function (CVDF) may vary with environment.  In this paper, we will compare the HIMF and CVDF 
for our six loose groups with those of the galaxy population in general and in other specific environments.  

This is the second of two papers concerning our survey.  In the first paper \citep[][hereafter Paper~I]{pis07}, 
we described our selection criteria, survey parameters, observations, data reduction, and the 
survey goals.  These will be briefly summarized in this paper.  We have already used our data to 
place constraints on the amount of intra-group \HI\ clouds that may be analogous to the high-
velocity clouds \citep[HVCs;][]{WvW97} seen around the Milky Way.  Namely, that any such HVC analogs 
must reside within 90 kpc of galaxies and have a total \mhi$\lesssim$10$^8$\msun \citep[][Paper~I]{pis04}.  In this 
paper, we will discuss the reliability of our survey and data analysis, present our \HI\ data on the 
galaxies in the six groups, derive an HIMF and CVDF for the loose group environment, and 
discuss the implications for the effect of environment on galaxy formation.  
We summarize the sample selection in \S~\ref{sec:sample} and the observations and data reduction in \S~\ref{sec:obs}.
The results are presented in \S~\ref{sec:results} including a description of the reliability and completeness of the survey, 
how we measure the galaxy properties, and a summary of the group and galaxy properties.  Finally, we present the
HIMF in \S~\ref{sec:himf}, the CVDF in \S~\ref{sec:cvdf} and our conclusions in \S~\ref{sec:conclusions}.

\section{Sample Selection}
\label{sec:sample}

For this project, we identified six poor, loose groups of galaxies that are analogous to the Local 
Group.  Details are given in Paper~I, but the selection is summarized here.  Groups were selected
to be nearby (\vgsr\ $<$ 1000 \kms), but not confused with Galactic \HI\ emission (\vgsr\ $>$ 300 
\kms).  The groups only contain spiral or irregular galaxies separated, on average, by
a few hundred kiloparsecs with a total extent of $\sim$1 Mpc.  Because our observations were 
made with the Parkes radio telescope in Australia, we only chose groups below a declination of
0\arcdeg.  We selected five groups, LGG~93, LGG~106, LGG~180, LGG~293, and LGG~478, 
from the Lyon Groups of Galaxies (LGG) catalog of \citet{garcia93}
and a sixth group from the HICAT group catalog of \citet{ste05}.  Distances are corrected using the
multiattractor velocity flow model of \citet[][K.L. Masters 2010, private communication]{mas05}, and
assuming $H_0$ = 72 \kms\ Mpc$^{-1}$ \citep{spe03}.  The measured properties of the groups are discussed
in detail in \S~\ref{sec:group}.  

\section{Observations \& Data Reduction}
\label{sec:obs}

We observed the entire extent of the six groups between 2001 October and 2003 June using 
the 20 cm multibeam instrument \citep{sta96} on the Parkes 64m radio telescope\footnote{The Parkes radio telescope is 
part of the Australia Telescope National Facility which is funded by the Commonwealth of Australia for 
operation as a National Facility managed by CSIRO.}.  Observations 
of the first two groups, LGG 93 and LGG 180, were made with an 8 MHz bandwidth and 1.65\kms\
channels using the inner seven beams of the multibeam, while all subsequent observations
were made using all 13 beams, a 16 MHz bandwidth, and 3.3\kms\ channels.  All groups were 
observed only at night to avoid solar interference.  Maps were made by scanning in a 
basket-weave pattern in right ascension and declination with consecutive scans being offset
to result in uniform coverage perpendicular to the scan direction.  Data were calibrated using 
periodic observations of flux calibrator Hydra A with a resulting accuracy of about 10\%.  All data
were reduced and gridded using the {\sc LIVEDATA} and {\sc GRIDZILLA} packages.  The 
1$\sigma$, 3.3 \kms\ rms noise in these cubes range from 5.5-7.0 mJy, corresponding to
a \mhi\ of 3.5-11$\times$10$^5$\msun and \nhi\ of 2.8-4.6$\times$10$^{16}$cm$^{-2}$ 
depending on the group.   For a 5$\sigma$ detection of a source with a 30 \kms\ linewidth, we have
a \mhi\ limit of (0.5-2)$\times$10$^7$\msun, and a \nhi\ limit of (4.2-6.9)$\times$10$^{17}$\cmsq.  Although
it is worth noting that our reduction technique will subtract out sources that are larger than a few beams
across.  

The final cubes were searched by three groups of authors:  DJP, DGB, and BKG and VAK in tandem.
Our final list of putative sources included those that were identified by at least two of the three groups of
authors.  In addition, our identification of fake sources added to our
cubes by M. Zwaan allowed us to assess the completeness of our survey, as discussed in Paper~I.

We used the Australia Telescope Compact Array (ATCA)\footnote{The Australia Telescope is 
part of the Australia Telescope National Facility which is funded by the Commonwealth of Australia for 
operation as a National Facility managed by CSIRO.} 
to confirm the reality of the sources identified in our Parkes data.  We observed 105 of the 112 
Parkes detections in and behind the six groups in our sample.  The remaining seven galaxies 
were all behind the groups and previously detected by HIPASS \citep{mey04}.  
Of the 105 sources, 15 had previously been observed with the ATCA or VLA with similar resolution and 
equal or better sensitivity than our original observations.  Data cubes for IC~1959, ESO~348-G9, IC~5332,
ESO~249-G35, ESO~249-G36, IC~2000, NGC~5084, and ESO~576-G40 came from project C~934 courtesy
of Emma Ryan-Weber.  Data for NGC~2997 comes from the GMRT and ATCA and were discussed in
detail in \citet{hess09}.  The remaining archival data was taken from the ATCA or VLA archives and re-reduced.
The galaxies and projects with archival ATCA data are as follows:  ESO~347-G29 (C073), NGC~1433 (C305), 
NGC~1448 (C295, C419), IC~1986 (C631, C942), UGCA~289 (C1046), NGC~5068 (C892), and 
ESO~575-G61 (C894).  Those with VLA data are DDO~146 (AD474) and UGCA~320 (AC320).  

We observed the remaining 90 sources
between 2002 October and 2005 March using a compact configuration with baselines shorter than
750m yielding beams of $\sim$1$\arcmin$-2$\arcmin$, with the exception of galaxies towards the 
equatorial group LGG~293.  For those sources we used the H214C configuration, which utilizes the 
north spur of the ATCA and has a maximum baseline of 214m producing a beam of 
$\sim$2$\arcmin$-3$\arcmin$.  Sources were observed to at least the same point-source 
sensitivity as the Parkes observations, $\sim$4 mJy beam$^{-1}$.  

More details on all the observations, data reduction, and source-finding can be found in Paper~I.

\section{Survey Results}
\label{sec:results}

\subsection{Reliability \& Completeness}
\label{sec:rel}

Our ATCA observations served two purposes.  The primary purpose was to establish the 
reality of the putative Parkes detections, while the secondary purpose was to identify any
\HI-rich galaxies that were confused at the Parkes resolution.  Our original and archival
interferometer observations confirmed the reality of 106 Parkes detections and revealed an
additional four dwarf galaxies behind the target groups that were confused in the original Parkes 
data.  As such our current sample of 110 \HI-rich galaxies is 100\% reliable.  

The completeness, which is a necessary measure of a survey if one wishes to construct mass 
functions, is far more complicated and is discussed in detail in Paper~I.  Figure 3 of Paper~I  shows
the completeness as a function of the linewidth and integrated signal-to-noise of the source based on
the fake sources that were inserted into our Parkes data cubes.  In Paper~I, we showed that the 
completeness we inferred was well-described by the completeness function for HIPASS from
\citet{zwaan04} after adjusting it for the different channel widths and noise levels of the two surveys.  As in
Paper~I, we use that scaled completeness function in this paper in order to derive the HIMF
and CVDF.  

\subsection{Measuring Galaxy Properties}
\label{sec:measure}

\begin{figure}
\includegraphics[width=0.9\textwidth]{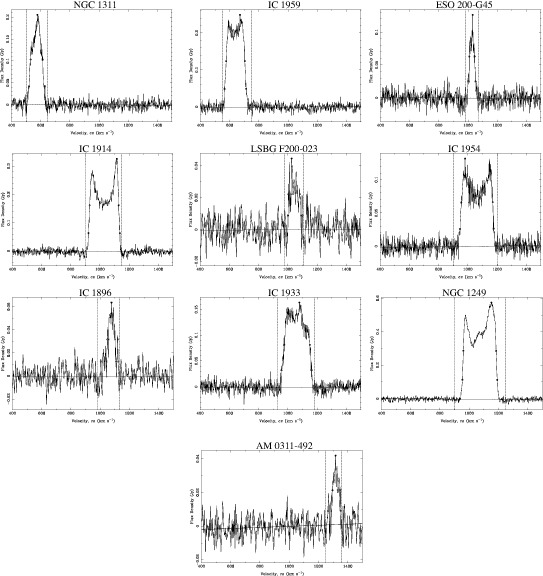}
\caption{Parkes \HI\ spectra of the confirmed detections in LGG~93.  The vertical dashed
lines indicate the range of velocites over which the profile properties were measured.  The
filled circles indicate the peak of the profile; the open circles the maximum 20\% and 50\% 
velocity widths; the $\times$'s mark the minimum 20\% and 50\% widths.  The nearly 
horizontal solid line indicates the baseline fit to the spectrum.  \label{fig:lgg93_spec}}
\end{figure}

\begin{figure}
\includegraphics[width=0.9\textwidth]{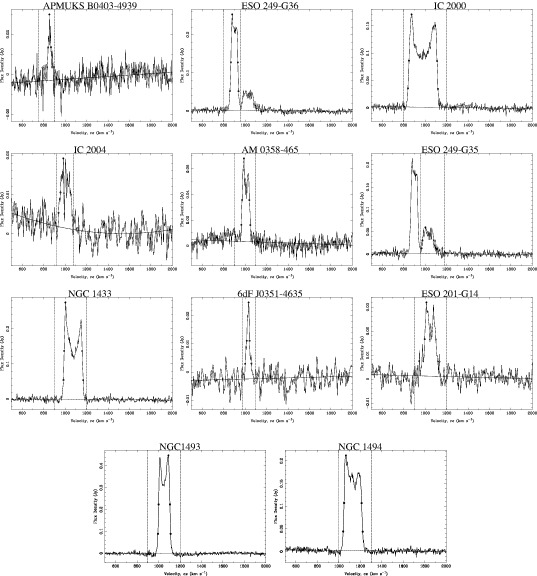}
\caption{Same as Figure~\ref{fig:lgg93_spec} but for LGG~106 group galaxies.
\label{fig:lgg106_spec}}
\end{figure}

\begin{figure}
\includegraphics[width=0.9\textwidth]{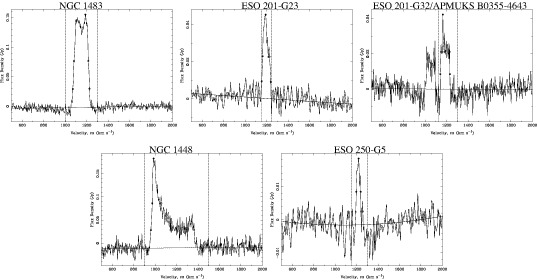}
\caption{Same as Figure~\ref{fig:lgg93_spec} but for the remaining LGG~106 group galaxies.
\label{fig:lgg106_spec2}}
\end{figure}

\begin{figure}
\includegraphics[width=0.9\textwidth]{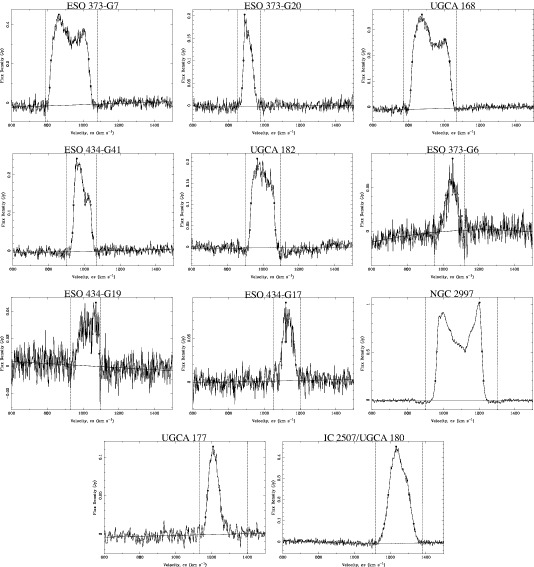}
\caption{Same as Figure~\ref{fig:lgg93_spec} but for LGG~180 group galaxies.
\label{fig:lgg180_spec}}
\end{figure}

\begin{figure}
\includegraphics[width=0.9\textwidth]{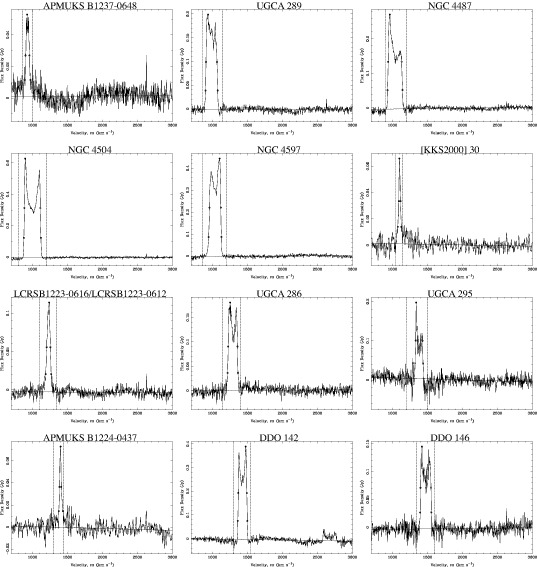}
\caption{Same as Figure~\ref{fig:lgg93_spec} but for LGG~293 group galaxies.
\label{fig:lgg293_spec}}
\end{figure}

\begin{figure}
\includegraphics[width=0.9\textwidth]{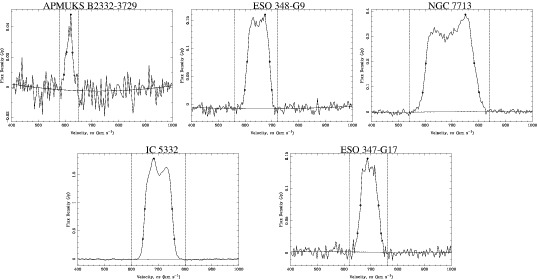}
\caption{Same as Figure~\ref{fig:lgg93_spec} but for LGG~478 group galaxies.
\label{fig:lgg478_spec}}
\end{figure}

\begin{figure}
\includegraphics[width=0.9\textwidth]{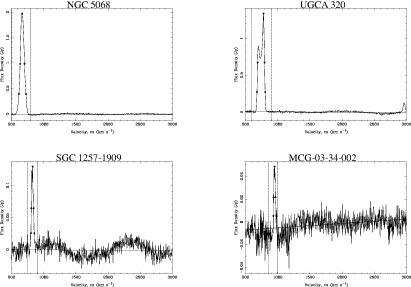}
\caption{Same as Figure~\ref{fig:lgg93_spec} but for HIPASS group galaxies.
\label{fig:hgrp3_spec}}
\end{figure}

\begin{figure}
\includegraphics[width=0.9\textwidth]{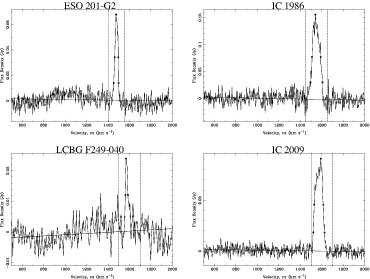}
\caption{Same as Figure~\ref{fig:lgg93_spec} but for galaxies behind LGG~106.
\label{fig:lgg106_bg_spec}}
\end{figure}

\begin{figure}
\includegraphics[width=0.9\textwidth]{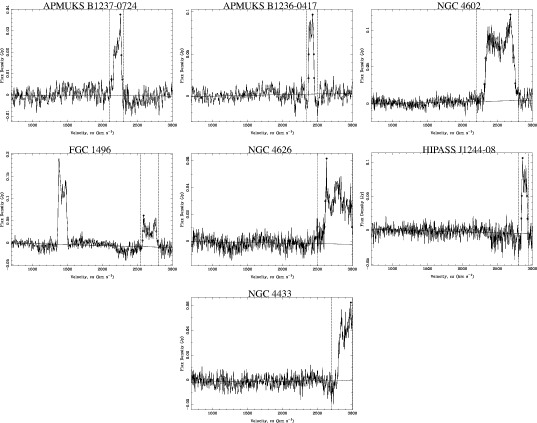}
\caption{Same as Figure~\ref{fig:lgg93_spec} but for galaxies behind LGG~293.
\label{fig:lgg293_bg_spec}}
\end{figure}

\begin{figure}
\includegraphics[width=0.9\textwidth]{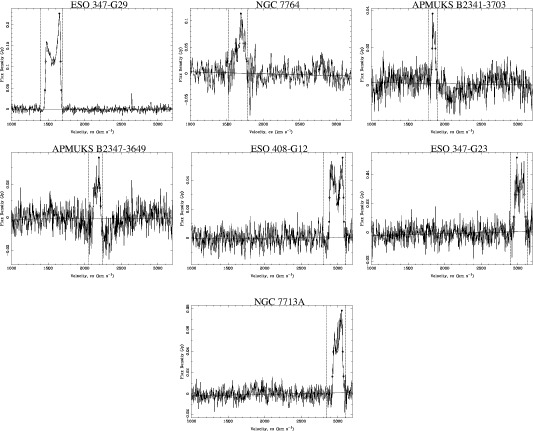}
\caption{Same as Figure~\ref{fig:lgg93_spec} but for galaxies behind LGG~478.
\label{fig:lgg478_bg_spec}}
\end{figure}

\begin{figure}
\includegraphics[width=0.9\textwidth]{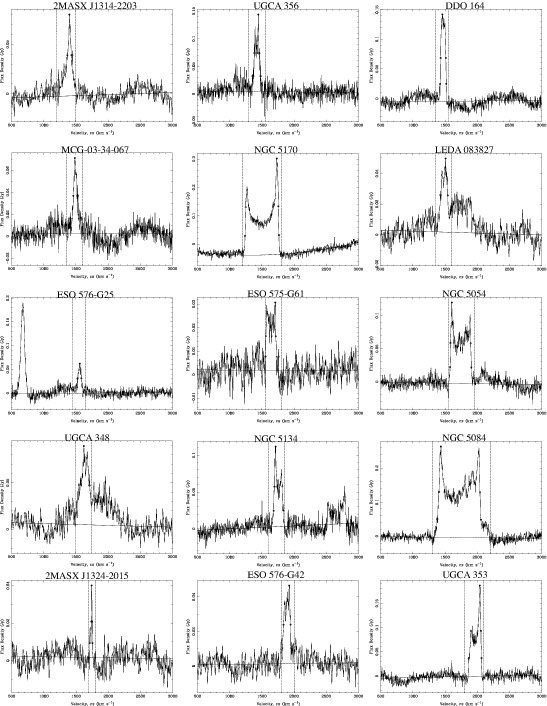}
\caption{Same as Figure~\ref{fig:lgg93_spec} but for galaxies behind the HIPASS group.
\label{fig:hgrp3_bg_spec}}
\end{figure}

\begin{figure}
\includegraphics[width=0.9\textwidth]{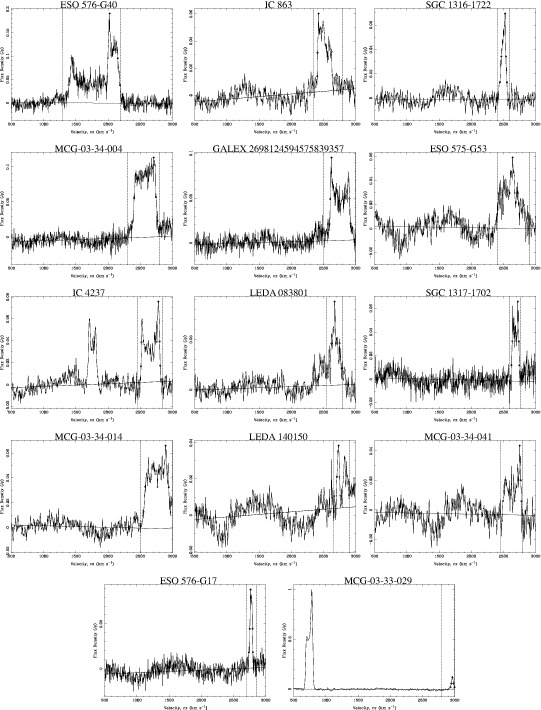}
\caption{Same as Figure~\ref{fig:lgg93_spec} but for the remaining 
galaxies behind the HIPASS group.
\label{fig:hgrp3_bg2_spec}}
\end{figure}

Basic source parameters, such as position, recessional velocity,
integrated \HI\ flux, and linewidth were measured from the Parkes data 
using the {\sc MBSPECT} task in {\sc MIRIAD}.  While searching the cubes, 
initial positions and velocities were determined for each source.  Using 
{\sc MBSPECT}, we inspected the cube at these positions to determine a velocity 
range to fit a first or  second order spectral baseline (for any residual shape
not removed in the reduction process).  {\sc MBSPECT} takes an input
position and velocity and a range of velocities to fit a baseline and
measure the \HI\ profile.  It then creates a moment map over the
latter range and fits a  Gaussian to determine the central position of
the source.  {\sc MBSPECT} then forms the spectrum in a 28$\arcmin
\times$28$\arcmin$~ box centered at this position.  For weak sources,
the spectra are Hanning smoothed.  The resulting spectrum is then
robustly integrated and the velocity width at the first and last
crossings of 20\% of the peak flux are identified.   Typically, we
chose the maximum width here, but when this was corrupted by noise
spikes, the minimum or an average of the minimum and maximum width 
was chosen instead.  For the two galaxies behind LGG~293 that are at the
edge of the observed band, we can only achieve lower limits or highly uncertain
estimates of the linewidth and integrated fluxes.  The Parkes \HI\ spectra of all detections, including indications of the peak flux
as well as the 50\% and 20\% crossings, are shown in Figures~\ref{fig:lgg93_spec}--\ref{fig:hgrp3_bg2_spec}.  
Where there were multiple galaxies within the 
Parkes beamwidth, we used the ATCA data to measure the galaxy properties.  
This was done in the same fashion, except the position was fixed on the known
location of the emission and the box in which the line was measured was 
defined to tightly enclose the visible emission.  

We can estimate the errors for our measured parameters by comparing
real and measured values for the detected fake sources (see Paper~I for
details on the fake sources) and by comparing
our measured paramters with those in HICAT \citep{mey04}.  For the 
fake sources, our position uncertainty is 2$\arcmin$.  For the remainder 
of the parameters, after discarding pathological outliers (more than 10$\sigma$, discussed below),
$V$, $W_{20}$, and $S_{int}$ show a scatter of 2 \kms, 4 \kms, and 0.3 Jy \kms, respectively.
This is much better than the errors for  HICAT \citep{zwaan04}, even though 
our fake sources also tend to be fainter, on average, than Zwaan \etal's.
For those sources with large discrepancies, they are mostly faint sources where noise
spikes are artificially broadening the velocity width measurements or sources 
with poor baseline fits.  

For the brighter galaxies in our sample, we can also compare our
measured parameters with those from HICAT \citep{mey04}.  There are
a total of 65 sources in our survey that are also in HICAT.  The positional
uncertainty is 2$\arcmin$ and the robust standard deviation, after discarding those sources
more than 10$\sigma$ from the mean, of $V$, $W_{20}$, and $S_{int}$
are 1 \kms, 4 \kms, and 0.9 Jy \kms.  For $W_{20}$, the HICAT widths are systematically
larger by $\sim$15 \kms\ due to the coarser velocity resolution of HIPASS.  The outliers 
for the HICAT sources are mostly faint sources or those with bad baselines, as for the fake
sources, but there was at least one source that was partly confused with another galaxy.  
Overall, we feel confident that we can accurately measure the properties of the galaxies 
from their \HI\ spectra.  

Tables~\ref{tab:grpdet} and \ref{tab:bgdet} list the measured \HI\ 
properties of the confirmed group and background galaxies.  

\subsection{Group Properties}
\label{sec:group}

\begin{figure}
\includegraphics[width=0.9\textwidth]{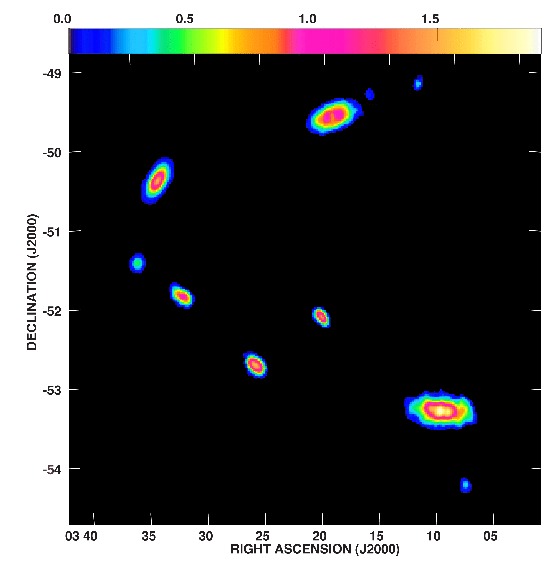}
\caption{Total \HI\ intensity maps of group galaxies in LGG~93 on the same intensity scale
(in units of 10$^{21}$cm$^{-2}$).  The galaxies have been placed at their correct locations, but
have been scaled up in size by a factor of five.  
\label{fig:lgg93_m0}}
\end{figure}

\begin{figure}
\includegraphics[width=0.9\textwidth]{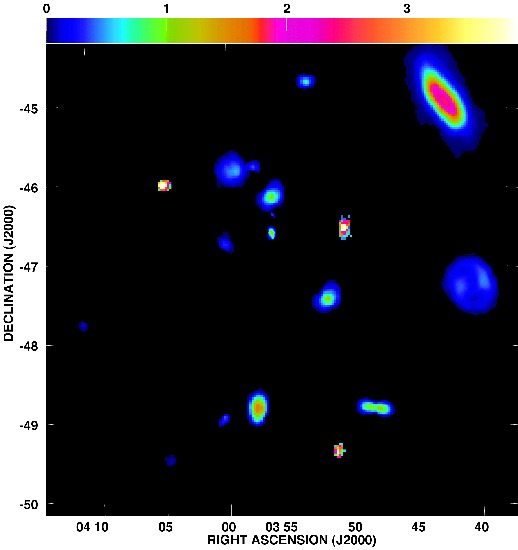}
\caption{Same as Figure~\ref{fig:lgg93_m0}, but for LGG~106.
\label{fig:lgg106_m0}}
\end{figure}

\begin{figure}
\includegraphics[width=0.9\textwidth]{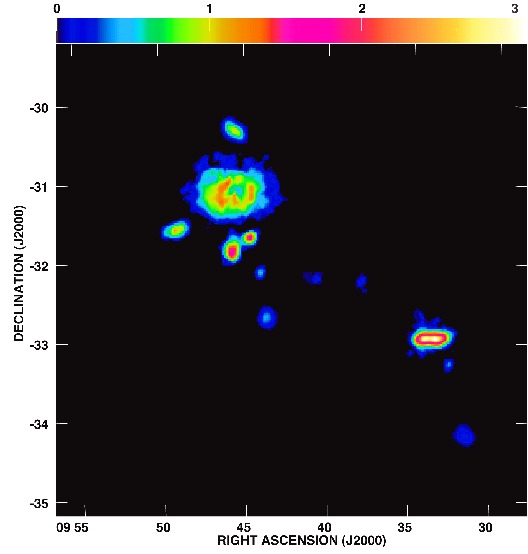}
\caption{Same as Figure~\ref{fig:lgg93_m0}, but for LGG~180.
\label{fig:lgg180_m0}}
\end{figure}

\begin{figure}
\includegraphics[width=0.9\textwidth]{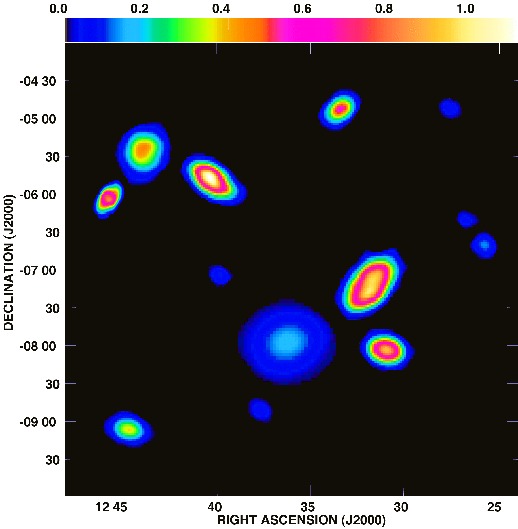}
\caption{Same as Figure~\ref{fig:lgg93_m0}, but for LGG~293.
\label{fig:lgg293_m0}}
\end{figure}

\begin{figure}
\includegraphics[width=0.9\textwidth]{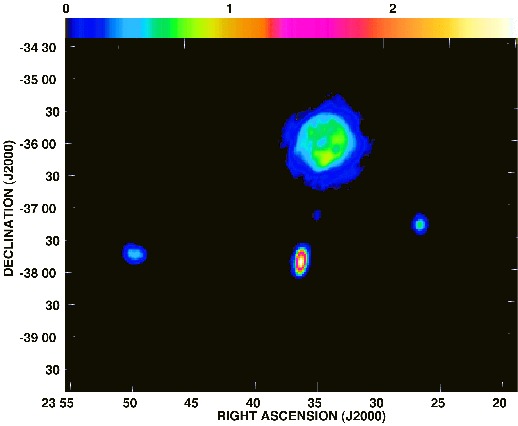}
\caption{Same as Figure~\ref{fig:lgg93_m0}, but for LGG~478.
\label{fig:lgg478_m0}}
\end{figure}

\begin{figure}
\includegraphics[width=0.9\textwidth]{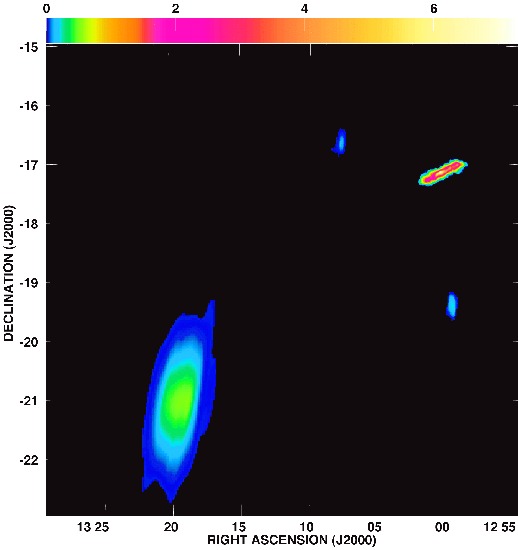}
\caption{Same as Figure~\ref{fig:lgg93_m0}, but for the HIPASS group.
\label{fig:hgrp3_m0}}
\end{figure}

A total of 31 galaxies detected in our \HI\ survey were previously identified by \citet{garcia93} to be associated with the six groups
we observed, but how many of the 110 \HI-rich galaxies we detected are also associated with the groups?  To answer this question, we used an
iterative process.  Starting with the optical group velocity and velocity dispersion, we identified those \HI-detected galaxies within 
three times the velocity dispersion.  The mean recession velocity and velocity dispersion is then recalculated and new group members 
are identified within 3$\sigma$ of the central velocity.  The process is repeated until both values have changed by less than 1\%.  The derived
values are listed in Table~\ref{tab:grpprop}.   We characterize the radial extent of each group in a few different ways.  The diameter of each group 
is taken to be twice the projected separation of the most distant galaxy from the group center.  As we assumed that the groups did not extend beyond
our survey area, the diameters tend to be comparable to the diagonal across the survey area.  We also calculated the mean projected separation
between group galaxies, and the projected radial dispersion of the group galaxies from the optically-defined group center.  Using this approach, we 
identified a total of 61 group galaxies in the six groups.  Overall, our survey roughly doubled the number of group members found by \citet{garcia93}.  
To illustrate the relative locations of the group galaxies, Figures~\ref{fig:lgg93_m0}--\ref{fig:hgrp3_m0} show the \HI\ total intensity 
(moment 0) maps of each group galaxy in its proper location, but scaled up in size by a factor of five.  

For each group, we calculated a mass, first assuming that they were virialized:
\begin{equation}
M_{vir}=\frac{3\pi (N-1)}{2G}\frac{\Sigma v^2_i}{\Sigma 1/R_{ij}}
\end{equation}
from \citet{heisler85} where R$_{ij}$ is the projected separation between a pair of galaxies and v$_i$ is the velocity difference between the galaxy and the group velocity, 
and N is the number of group members.  If the groups are not virialized, then we can use the projected-mass estimator to infer their masses:
\begin{equation}
M_{pm}=\frac{32 \Sigma v^2_i R_i}{\pi G (N-1)}
\end{equation}
where isotropic orbits are assumed and R$_i$ is the separation between the galaxy and the group center \citep{heisler85}.  For either mass, we are able to calculate the 
zero-velocity surface for the bound group:
\begin{equation}
R_0 = (\frac{8 G T^2 M}{\pi^2})^{1/3}
\end{equation}
where $M$ is the mass and $T$ is the age of the group \citep{sandage86}, taken to be 13.7 Gyr, the age of the universe \citep{spe03}.  

In Table~\ref{tab:grpprop}, we list the derived group properties for each of the six groups.  Our selection of the groups as analogs to
the Local Group was based purely on the morphology of the galaxies--group members are only spirals and irregulars, the group members being
widely spaced, and the absence of detectable intragroup X-ray emission.  From Table~\ref{tab:grpprop}, we see that all of the groups are very 
similar to each other in terms of extent, morphology, and derived masses.  Their average velocity dispersion is 133$\pm$59 \kms; their
average diameter is 1.1$\pm$0.2 Mpc; and the average separation of the galaxies in the groups is 525$\pm$85 kpc.  For comparison, we used the 
measured distances and positions of Local Group galaxies with \mhi$\ge$10$^7$\msun\ to calculate a group diameter of $\sim$3.8 Mpc, a
mean galaxy separation of $\sim$1.3 Mpc, and a radial dispersion of 530 kpc \citep{mateo98,vdb00}.  \citet{vdb00} report that $\sigma_v$=61$\pm$8 \kms\
for the Local Group.  If the groups are all virialized, their median mass is 6$\times$10$^{12}$\msun\ with an average zero-velocity surface at 1.5$\pm$0.8 Mpc.  
If the groups are not virialized, then we derive a median mass of 1.3$\times$10$^{12}$\msun\
with an average zero-velocity surface of 2.1$\pm$0.8 Mpc.   The derived sizes and masses of these loose groups,
whether these systems are virialized or not, are all similar to the Local Group with M$_{vir}$=(2.3$\pm$0.6)$\times$10$^{12}$\msun\ 
and R$_0$=1.15$\pm$0.15 Mpc \citep{vdb00}.  As such, we remain confident that these groups are good analogs for the Local Group.

\subsection{Galaxy Properties}
\label{sec:galaxy}

\begin{figure}
\includegraphics[width=0.9\textwidth]{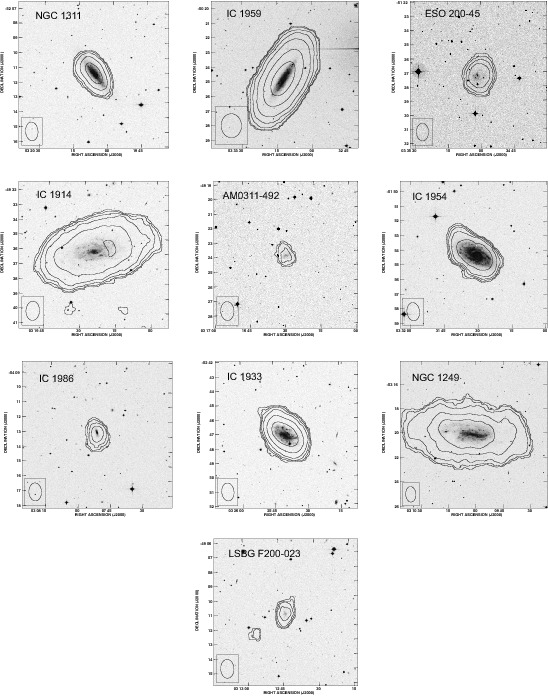}
\caption{ATCA total \HI\ intensity (moment 0) contours overlaid on second generation 
blue Digital Sky Survey greyscale images for LGG~93 group galaxies.  Contour levels are given in Table~\ref{tab:grpcubes}.  
The beam is shown as the boxed ellipse at the bottom of each image.
\label{fig:lgg93_opt}}
\end{figure}

\begin{figure}
\includegraphics[width=0.9\textwidth]{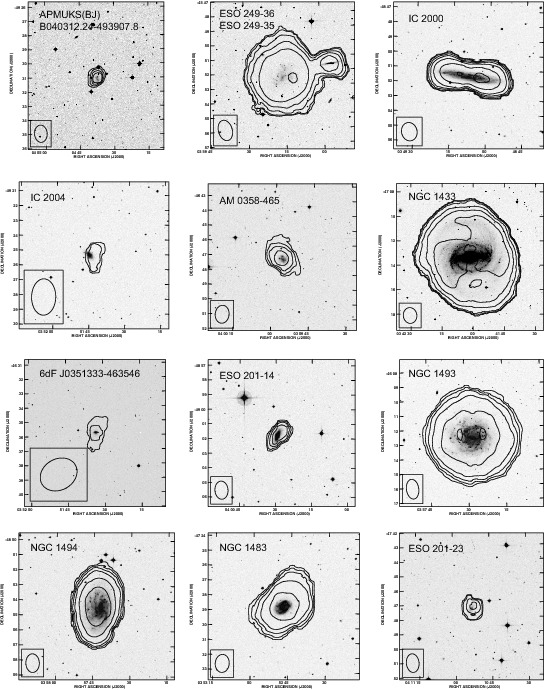}
\caption{Same as Figure~\ref{fig:lgg93_opt}, but for LGG~106 group galaxies.
\label{fig:lgg106_opt}}
\end{figure}

\begin{figure}
\includegraphics[width=0.9\textwidth]{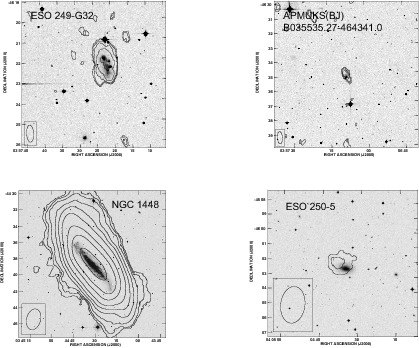}
\caption{Same as Figure~\ref{fig:lgg93_opt}, but for the remaining LGG~106 group galaxies.
\label{fig:lgg106_opt2}}
\end{figure}

\begin{figure}
\includegraphics[width=0.9\textwidth]{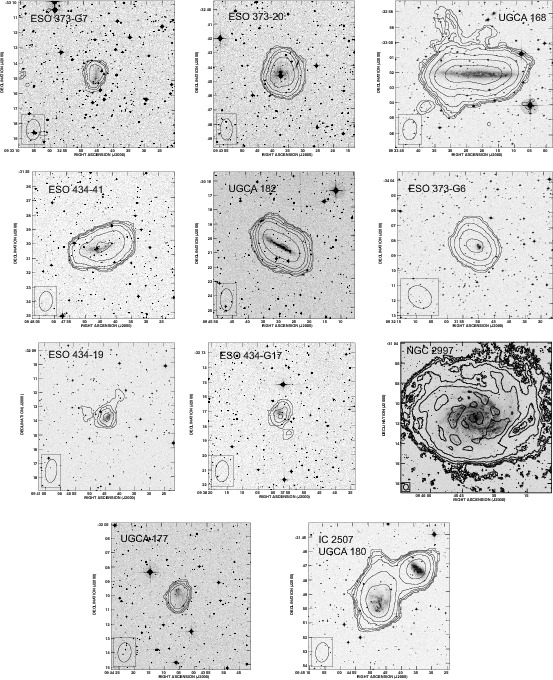}
\caption{Same as Figure~\ref{fig:lgg93_opt}, but for LGG~180 group galaxies.
\label{fig:lgg180_opt}}
\end{figure}

\begin{figure}
\includegraphics[width=0.9\textwidth]{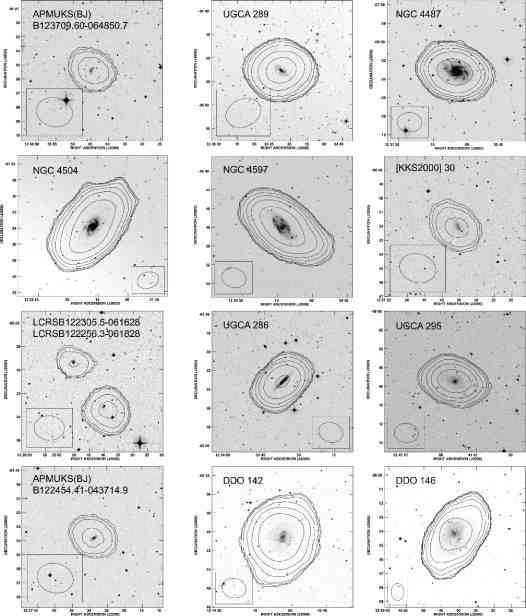}
\caption{Same as Figure~\ref{fig:lgg93_opt}, but for LGG~293 group galaxies.
\label{fig:lgg293_opt}}
\end{figure}

\begin{figure}
\includegraphics[width=0.9\textwidth]{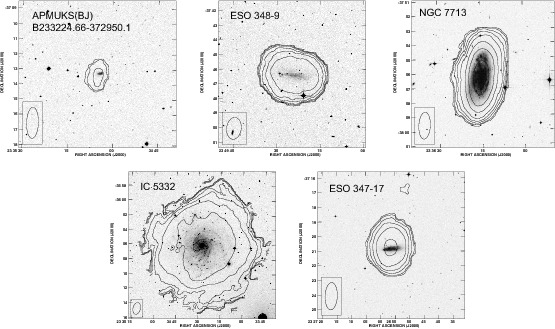}
\caption{Same as Figure~\ref{fig:lgg93_opt}, but for LGG~478 group galaxies.
\label{fig:lgg478_opt}}
\end{figure}

\begin{figure}
\includegraphics[width=0.9\textwidth]{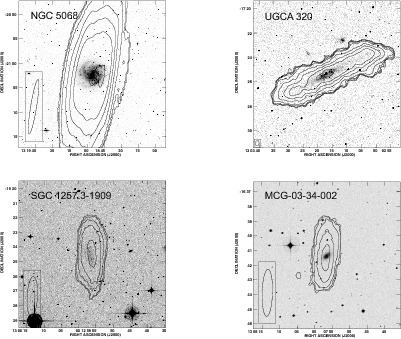}
\caption{Same as Figure~\ref{fig:lgg93_opt}, but for HIPASS group galaxies.
\label{fig:hgrp3_opt}}
\end{figure}

\begin{figure}
\includegraphics[width=0.9\textwidth]{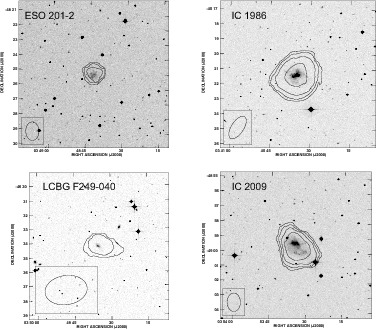}
\caption{Same as Figure~\ref{fig:lgg93_opt}, but for galaxies behind LGG~106.  The contour
levels are listed in Table~\ref{tab:bgcubes}.
\label{fig:lgg106_bg_opt}}
\end{figure}

\begin{figure}
\includegraphics[width=0.9\textwidth]{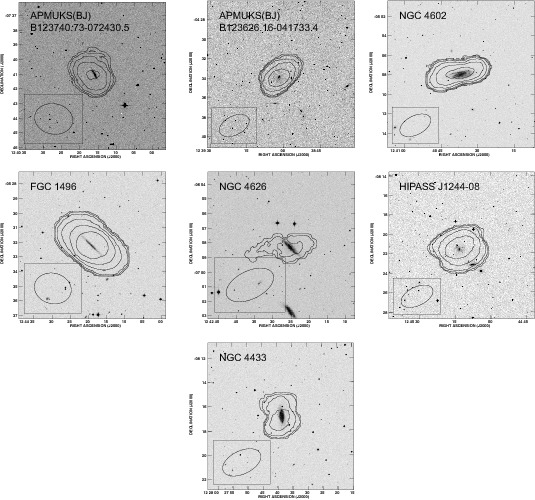}
\caption{Same as Figure~\ref{fig:lgg106_bg_opt}, but for galaxies behind LGG~293.
\label{fig:lgg293_bg_opt}}
\end{figure}

\begin{figure}
\includegraphics[width=0.9\textwidth]{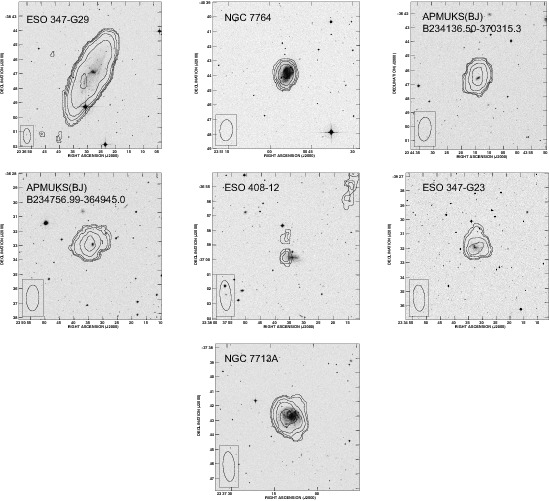}
\caption{Same as Figure~\ref{fig:lgg106_bg_opt}, but for galaxies behind LGG~478.
\label{fig:lgg478_bg_opt}}
\end{figure}

\begin{figure}
\includegraphics[width=0.9\textwidth]{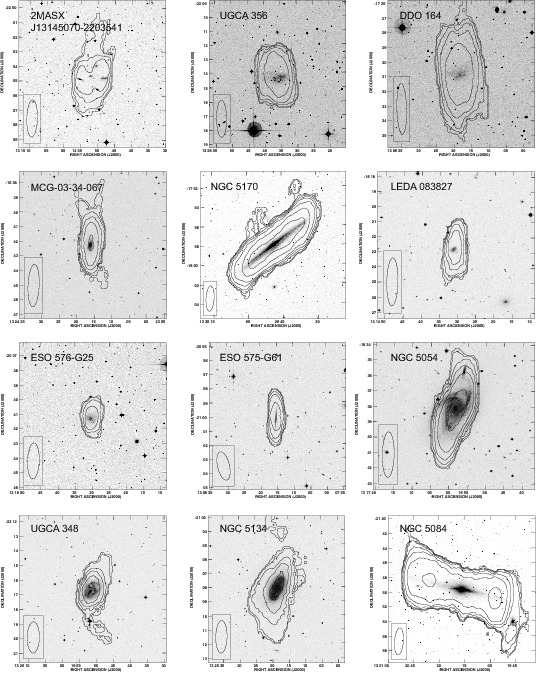}
\caption{Same as Figure~\ref{fig:lgg106_bg_opt}, but for galaxies behind the HIPASS group.
\label{fig:hgrp3_bg_opt}}
\end{figure}

\begin{figure}
\includegraphics[width=0.9\textwidth]{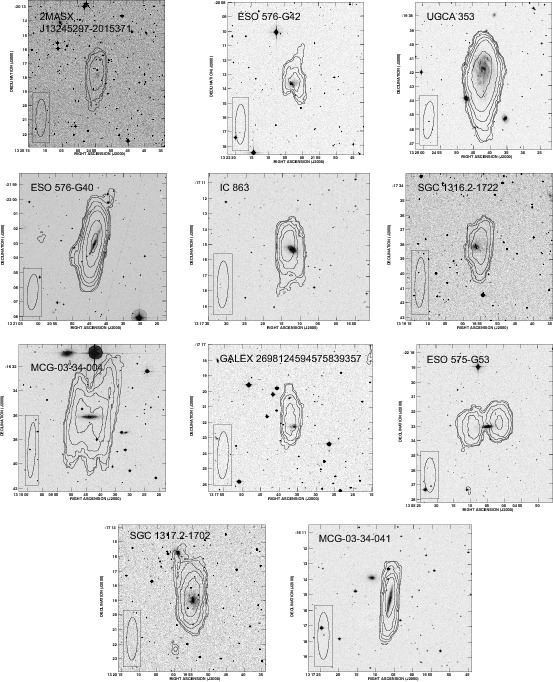}
\caption{Same as Figure~\ref{fig:lgg106_bg_opt}, but for the remaining galaxies behind 
the HIPASS group. \label{fig:hgrp3_bg2_opt}}
\end{figure}

Interferometer \HI\ total intensity (moment 0) maps overlaid on optical images for all group galaxies are 
shown in Figures~\ref{fig:lgg93_opt}--\ref{fig:hgrp3_opt}.  \HI\ total intensity contours on optical maps 
for the background galaxies are shown in 
Figures~\ref{fig:lgg106_bg_opt}--\ref{fig:hgrp3_bg2_opt}.  The properties of the 
interferometer observations are listed in Tables~\ref{tab:grpcubes} and 
\ref{tab:bgcubes}.  For the remainder of this paper, we will limit our discussion to those
galaxies belonging to the targeted groups.  

\begin{figure}
\includegraphics[width=0.9\textwidth]{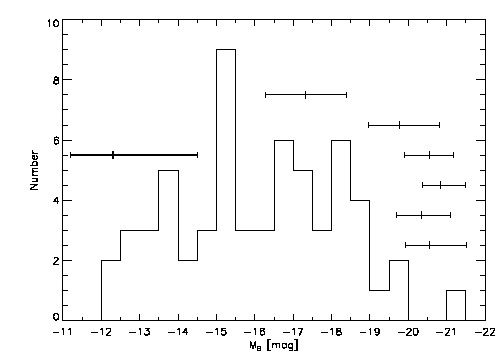}
\caption{A histogram showing the distribution of absolute B-band magnitudes for the group
galaxies.  The error bars indicate the 25th, 50th, and 75th percentile distribution of M$_B$ for 
UGC galaxies as reported by \citet{roberts94} and corrected for H$_0$=72 \kms\ Mpc$^{-1}$.  
From bottom to top they represent these values for E/S0, S0a/Sa, Sab/Sb, Sbc/Sc, Scd/Sd, and Sm/Im
galaxies.  The thick error bar indicates the same thing for Local Group dwarf galaxies with \HI\ detections from \citet{mateo98} and
\citet{kalirai10}.  The group galaxies detected in \HI\ are preferentially lower luminosity galaxies compared to
traditional galaxies on the Hubble sequence; our survey is primarily detecting dwarf galaxies.  \label{fig:mb}}
\end{figure}

\begin{figure}
\includegraphics[width=0.9\textwidth]{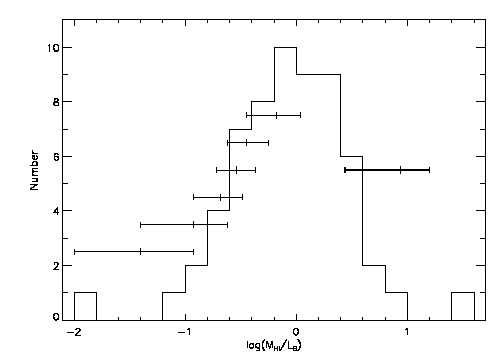}
\caption{A histogram of the \mhi/L$_B$ ratio for our group galaxies.  The error bars are as in Figure~\ref{fig:mb}, but for the \mhi/L$_B$ ratio.
Local Group dwarf galaxy data from \citet{mateo98}, \citet{grcevich09}, and \citet{kalirai10}.  
Approximately half of our group galaxies detected in \HI\ are more gas-rich than even Magellanic spirals and irregulars
\citep{roberts94}.  Again, our survey is primarily finding gas-rich dwarf galaxies. \label{fig:ml}}
\end{figure}

\begin{figure}
\includegraphics[width=0.9\textwidth]{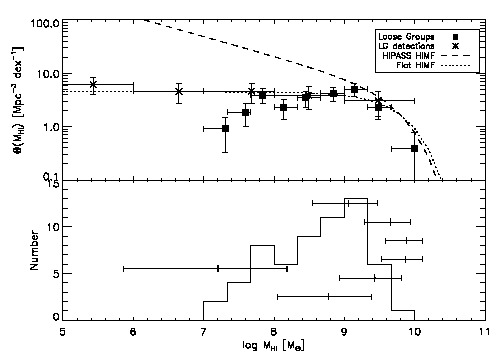}
\caption{Top:  The \HI\ mass function for our sample of six loose groups
(squares) as compared to the Local Group galaxies detect in \HI\ (asterisks).  
\mhi\ for Local Group galaxies comes from \citet{mateo98} and \citet{grcevich09}.  
The points are plotted at the mean \mhi\ for the galaxies in each bin.  The
horizontal extent of the error bars represents the bin size, while their 
vertical extent represents the Poisson noise.  The solid line represents a
flat Schechter function ($\alpha=-1.0$) roughly normalized to the Local
Group.  The dashed line is the HIPASS \HI\ mass function from \citet{zwaan05}
normalized to match our data points.  
Bottom:  The raw number of galaxies in each \mhi\ bin for our 
loose groups. The error bars are as in Figure~\ref{fig:mb}, but for \mhi.  Local Group dwarf galaxy data are
from \citet{mateo98} and \citet{grcevich09}.  It is clear from this
comparison that most of the galaxies have \mhi\ consistent with a late-type spiral galaxy or a dwarf irregular galaxy
\label{fig:himf}}
\end{figure}

The derived properties of the group galaxies are listed in Table~\ref{tab:grpgalprops}.  As shown in Figures~\ref{fig:mb}--\ref{fig:himf}, these 
galaxies span a wide range of \HI\ mass with \mhi$\sim$10$^{7-10}$\msun\ and luminosity, from M$_B$= -12.4-- -21.3 mag 
(L$_B$=1.4$\times$10$^7$\lsun--5.1$\times$10$^{10}$\lsun).  These figures clearly show that most of the galaxies detected in \HI\
in these groups span the full range of luminosities and \mhi\ from the median of Local Group \HI-rich dwarfs right through to the median of typical 
spiral galaxies.  The galaxies also have \HI-mass-to-light ratios that are generally consistent with those of late-type spiral galaxies or Local Group \HI-rich dwarf 
irregulars with \mhi/L$_B\sim$0.1--5 \msun/L$_\odot$ with few exceptions.  Those exceptions are two extremely gas-rich galaxies:
ESO~348-G9 and ESO~373-G7 with \mhi/L$_B$ of 9.27 and 26 \msun/L$_\odot$, respectively, and one relatively
gas-poor galaxy, ESO~250-G5, a lenticular with  \mhi/L$_B$ of 0.01 \msun/L$_\odot$ that has a remarkably low 
\mhi$\sim$10$^7$\msun.  

\section{\HI\ Mass Function}
\label{sec:himf}

Our primary goal in this paper is to determine how the mass function of galaxies in the low density group
environment compares to that in the field in general and in denser environments.  We begin by examining the HIMF for loose groups.

To construct the HIMF, we followed the same bivariate stepwise maximum likelihood method described by \citet{zwaan03}, that accounts for the survey completeness as
a function of linewidth and integrated flux.  For the HIMF we placed each galaxy in 0.3 dex wide \mhi\ bins with a weight based on the scaled HIPASS completeness function for 
that galaxy (described in Paper~I).  To convert to a volume density of galaxies, we took the 
total volume of the survey assuming that each loose group covered the entire survey area and had a depth equal to its diameter for a total survey volume of
7.81 Mpc$^3$ for the six groups.  The result is shown at the top of Figure~\ref{fig:himf}.  
For comparison, we also show the HIMF for those Local Group galaxies that have been detected in \HI, using data from \citet{mateo98} and \citet{grcevich09} for the
dwarf galaxies, \citet{staveley-smith03} for the LMC, \citet{stanimirovic99} for the SMC, and \citet{vdb00} for the Milky Way, M~31, and M~33.  For 
this HIMF, we assumed that the volume of the Local Group was the same as the average loose group.  Aside from the two lowest \mhi\ bins, where the
completeness is the most uncertain, there is very good agreement in the slope and normalization of the HIMF for the Local Group and our sample of loose
groups.  For comparison, we have also shown the HIMF from the HIPASS galaxies derived by \citet{zwaan05} as described by a Schechter function of the form:
\begin{equation}
\Theta (M) = \Theta^\star \ln(10) (\frac{M_{HI}}{M^\star})^{\alpha+1} \exp(-\frac{M_{HI}}{M^\star})
\end{equation}
where M$^\star$ is the mass where the function transitions to a low-mass power-law with slope $\alpha$, and $\Theta^\star$ is the normalization.  For HIPASS, 
log \mhi$^\star$=9.8, $\Theta^\star$=0.006 Mpc$^{-3}$ dex$^{-1}$, and $\alpha$=-1.37 \citep{zwaan05}.  We also show an identical
HIMF with a faint end slope of $\alpha$=-1.0.  In this figure, these two HIMFs have been 
renormalized to approximately match the loose group and Local 
Group mass functions.  
While not shown on the figure, the
recent HIMF from ALFALFA \citep{martin10} is based on over twice as many galaxies and has a wider mass range than the HIPASS HIMF \citep{zwaan05}, although most of its 
low-mass sources are in the high density Virgo cluster.  Nevertheless, the resulting fit is not dramatically different with 
log \mhi$^\star$=9.96, $\Theta^\star$=0.0048 Mpc$^{-3}$ dex$^{-1}$, and $\alpha$=-1.33.  

There have been many recent measurements of how the low-mass slope, $\alpha$, of the HIMF varies with the local galaxy density.  \citet{zwaan05} calculated the HIMF
for HIPASS galaxies in different density regions and found that as the local density decreased, the slope became flatter.  \citet{zwaan05}, however, used the HIPASS catalog
to define the local galaxy density, so \HI\ observations of galaxies in specific environments are needed to provide independent confirmation of these results.  Such observations
generally support the conclusions of \citet{zwaan05}.  \citet{freeland09} assembled a HIMF for five groups, four of which lack 
X-rays and are spiral-rich, and found a flat, $\alpha$=-1.0, HIMF. Kova{\v c} \etal\ (2005, 2009) found a similar flat slope, $\alpha$= -1.07, in the low density Canes Venatici group, as
did \citet{verheijen01} for the low density Ursa Major cluster.  In higher density groups, that contain more early-type galaxies and X-ray emission, the results are more varied.  
\citet{kilborn09} found a declining low-mass slope, $\alpha$=0.0, while \citet{stierwalt09} found a steep low-mass slope close to that found by \citet{zwaan05}, $\alpha$= -1.41.  
In higher density clusters, the low-\mhi\ slope also tends to be steeper \citep{gavazzi05,gavazzi06}, however this may not hold in the centers of clusters where there is a lack of 
\HI-rich galaxies \citep[cf.][]{davies04,springob05}.    In light of these past results, our HIMF for the low density environments of the  Local Group and our sample of six analogous 
loose groups is consistent with a flattening HIMF in lower density environments.  This behavior is mimicked by the optical luminosity functions with flatter slopes being 
found in lower density environments \citep{tully02}, although \citet{croton05} found little variation in the faint-end slope with environment.   

\citet{zwaan05} provide two possible explanations for the flattening of the low-mass slope of the HIMF in low density environments.  The first is that since the star formation rate and
the specific star formation rate are enhanced in lower density environments, then the \HI\ in group galaxies will be consumed faster than in cluster galaxies, particularly at low
\mhi.  Enhanced merging of low-\mhi\ galaxies could also provide a surplus of high-\mhi\ galaxies in low density environments.  Both of these processes could lead to a flattening of the 
HIMF.  In the higher density clusters and groups that contain a hot, dense IGM, ram pressure stripping, as is seen in the Virgo cluster \citep{chung09}, could shift galaxies from 
higher \mhi\ to lower \mhi, causing the HIMF to steepen in these environments.  While the densities are low, two of our groups, LGG~93 and LGG~180, have \HI\ deficiencies of 
0.2$\pm$0.11 and 0.1$\pm$0.05 based on HIPASS data \citep{sengupta06}.  Even without an X-ray bright IGM, it is possible that these groups could have a cool, dense IGM 
\citep[e.g.][]{freeland11} that could cause ram pressure stripping.  Alternatively, stripping from tidal interactions could be occurring in at least these two groups, although neither form of 
stripping would explain the flat HIMF for our groups.  The variations of $\alpha\ $ between groups and clusters of similar density may be due to other effects.  For example, in order to 
sustain star formation beyond the next gigayear, we know that the Milky Way needs to accrete more gas \citep[e.g.][]{peek09}.  There have been many processes proposed to halt the 
accretion of cold gas onto galaxies, and, hence, halt star formation.  These include shock-heating the accreting gas to the virial temperature of the halo \citep{cattaneo06}, or heating 
from AGN, supernovae-driven winds, and/or star formation \citep{hopkins06}.  Finally, \citet{zwaan05} suggested that the flat slope of the HIMF in low density regions can be 
explained by the halo occupation model of \citet{mo04}, such that the late-type galaxies that dominate groups also reside in halos with a flat low mass slope.  Simulations of galaxy 
formation are just starting to become sophisticated enough to predict how \mhi\ varies with galaxy properties and environment \citep{duffy11}, so, in order to more directly compare 
with simulations, we construct the CVDF for loose groups.  

\section{Circular Velocity Distribution Function}
\label{sec:cvdf}

The CVDF uses the measured or inferred circular rotation velocity of a galaxy as a proxy for the mass of the dark matter halo within which the galaxy resides.  V$_{circ}$ is a
robust measure of the total dark matter mass in simulations and it can be compared with observations at least for those halos that host a
luminous galaxy.  An \HI\ survey, like our own, not only can identify faint, gas-rich galaxies but it provides a measure of the dynamical mass of the galaxy through the \HI\ 
linewidth.   Such surveys, however, are unlikely to detect the lowest mass galaxies, e.g. the dwarf spheroidals, that tend to lack detectable amounts of \HI\ \citep{grcevich09}.  

To construct a
CVDF for the six loose groups in this study, we start with the measured $W_{20}$ from the integrated \HI\ spectrum.  Following the procedure of \citet{meyer08}, we first correct 
$W_{20}$ for instrumental broadening using W$_{20,s}$=W$_{20}$-0.55R, where R is the spectral resolution (either 1.65 or 3.3 \kms\ for the Parkes data or in 
Table~\ref{tab:grpcubes} for ATCA data).  We then correct the linewidth for turbulent broadening following the prescription of \citet{tully85}:
\begin{equation}
W^2_R=W^2_{20,s}-W^2_t-2W_{20,s}W_t[1-e^{-(W_{20}/W_c)^2}]-2W^2_te^{-(W_{20}/W_c)^2}
\end{equation}
using $W_c$, the transition between single- and double-peaked profiles, of 120 \kms, and $W_t$, the turbulence correction, of 22 \kms to obtain the full rotation amplitude, 
W$_R$.  Finally, we apply an inclination correction to each linewidth and divide by two to get V$_{rot}$ as shown in Figure~\ref{fig:cvdf}.  We assume that V$_{rot}$ is 
equal to V$_{circ}$ for the associated dark matter halos of all of our group galaxies \citep{klypin99}.  Note that this approach assumes that all of the group galaxies are 
rotating and are not supported by random motions.  

\begin{figure}
\includegraphics[width=0.9\textwidth]{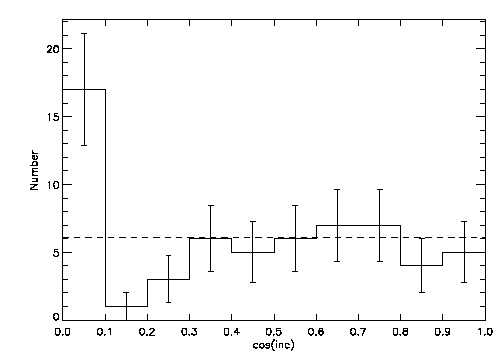}
\caption{The distribution of group galaxy inclinations.  If randomly selected, the cosine of the 
inclination should be a flat distribution.  The distribution deviates from a random distribution at
high inclinations. \label{fig:inc}}
\end{figure}

\begin{figure}
\includegraphics[width=0.9\textwidth]{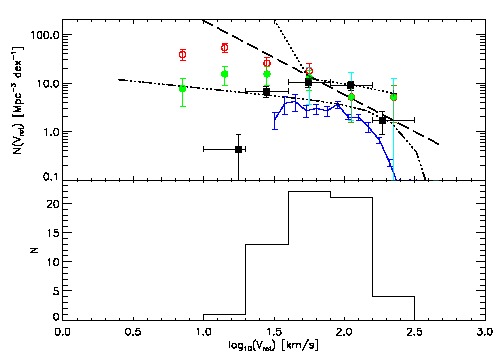}
\caption{Top:  The circular velocity distribution function (CVDF) 
for the Local Group (circles) and loose groups (squares).  The loose group
data is plotted at the mean V$_{rot}$ for each bin.  The filled
circles are the CVDF for the Local Group derived only for galaxies with
\HI\ detections.  The open circles include Local Group galaxies that have 
dynamical data from stellar kinematics.  The data for the Local Group
data for dwarf galaxies come from \citet{mateo98,simon07,kalirai10,walker09,geha09}.  
Data for the LMC come from \citet{kim98}, the SMC \citet{stanimirovic99}, M33 \citet{corbelli97}, while the 
Milky Way and M31 data are from \citet{vdb00}.  The solid line with error bars is the
CVDF for HIPASS detections from \citet{zwaan10}.  The dashed line represents the 
CVDF for cluster galaxies from \citet{desai04}, while the dot-dash line is for field
galaxies from \citet{gonzalez00}.  Finally, the dotted line with the cyan error bars
is the CVDF construction from the Via Lactea II simulations \citep{diemand08}.  All
CVDFs aside from those for the loose groups and the Local Group have been renormalized
to roughly match our data.  Bottom:  The raw number of galaxies in each V$_{rot}$ bin for
our loose groups.  \label{fig:cvdf}}
\end{figure}

For each galaxy, we took the inclination from 
HyperLEDA\footnote{\url{http://leda.univ-lyon1.fr/}} or used the same formula with the axial ratio from NED\footnote{\url{http://nedwww.ipac.caltech.edu/}}:
\begin{equation}
\sin^2 i = \frac{1-10^{-2 \log r25}}{1-10^{-2 \log r_o}}
\end{equation}
where $r25$ is the axial ratio and log r$_o$ = 0.38, appropriate for galaxies with Hubble types later than Sd \citep{paturel03}.  For one galaxy, APMUKS B1237-0648, this
method yielded sin i$>$1, so we chose $i$=90$\arcdeg$.  Very few of our galaxies were observed with sufficiently high resolution to derive a kinematic inclination, but for 
those galaxies that have high quality, high resolution data, NGC~1249, UGCA~168, NGC~2997, IC~5332, and UGCA~320, the kinematic inclinations agree to within 10$\arcdeg$
of the optical inclinations.  In Figure~\ref{fig:inc}, we plot the cosine of the inclination for all group galaxies.  While the
distribution is largely consistent with the disks being randomly distributed, there is a 2.6$\sigma$ excess of galaxies with cos i$<$0.1, a 5$\sigma$ deficit of galaxies with
cos i =0.1-0.2, and a 1.8$\sigma$ deficit of galaxies with cos i =0.2-0.3.  If all those extra galaxies with cos i$<$0.1 were redistributed to the other two bins, this
would result in an increase of derived V$_{rot}$ values of only 5\%.  Since the inclination also shows no correlation with either the integrated flux or the linewidth of the galaxy,
we have applied no correction to the inclinations.  To create the CVDF, the group galaxies were placed in bins of 0.3 dex width with a weighting based on the completeness
of the survey as a function of linewidth and integrated \HI\ flux.  This results in the solid points as shown in Figure~\ref{fig:cvdf}.  

For comparison, we have also created a CVDF for the Local Group galaxies both detected and undetected in \HI.  The linewidth data for the Local Group galaxies comes from 
a variety of sources:  for the Milky Way and M~31 the rotation velocity comes from \citet{vdb00}; M~33 from \citet{corbelli97}, the LMC from \citet{kim98}; the SMC from 
\citet{stanimirovic99}; and the rest of the dwarf galaxies from \citet{mateo98}, \citet{longmore82}, \citet{simon07}, \citet{martin07}, \citet{geha09}, \citet{walker09}, and 
\citet{kalirai10}.  For some of these galaxies, there are measurements of their V$_{rot}$ from stars or \HI.  For those galaxies without measured rotation, or which have 
velocity dispersions greater than the rotation velocity we assume isotropic orbits in an isothermal halo, so V$_{rot}$=$\sqrt{2}\sigma$.  In Figure~\ref{fig:cvdf} we plot the 
CVDF for all the Local Group galaxies and, separately, only for those which have been detected in \HI.   As for the HIMF, we assume that the average survey volume per 
group is equal to the volume of the Local Group.  Just like the HIMF, there is excellent agreement between the loose group CVDF and the Local Group CVDF for \HI-rich 
galaxies, except for the lowest V$_{rot}$ bin, where the completeness of the group survey is more poorly estimated.  

In addition to the Local Group, we compare the group CVDF with three other observed CVDFs and the results of simulations.  As for the HIMF, these CVDFs have been 
renormalized to approximately match the group CVDF.  The solid line with error bars is the HIPASS 
CVDF from \citet{zwaan10}
The ALFALFA velocity width function \citep{papastergis11} is almost identical to the HIPASS velocity width function \citep{zwaan10}.  Note that \citet{papastergis11} did not convert 
the velocity widths into rotation velocities, so it is not directly comparable to the other functions here, but does extend down to half the velocity width of the \citet{zwaan10} results 
and has almost five times the number of sources.  The dashed line is a cluster CVDF from \citet{desai04},
while the dot-dash line is for field galaxies from \citet{gonzalez00}.
Aside from the normalization, the slopes of the CVDF at low V$_{rot}$ are in good agreement between the loose groups, HIPASS, and the \citet{gonzalez00} field galaxies.  
In contrast, there is a deficit of low mass galaxies in loose groups as compared to the \citet{desai04} cluster sample.  In addition, if we compare to the CVDF from the 
Via Lactea II results \citep{diemand08},
we see a deficit of low mass galaxies in loose groups; this is the standard definition of the ``missing satellite" problem.  Note that 
the Via Lactea II simulation is a dark matter-only simulation of a Milky Way-sized halo and its sub-halos, so the resulting CVDF can not be directly compared to larger halos.  
The loose groups and Local Group, however, are not significantly more massive than the parent halo in these simulations.  

Examining the CVDF as a function of galaxy density from low density groups through the field to the cluster environment, we do not see any significant differences except when
we compare to the highest density cluster environment.  While \citet{zwaan10} did not look at the CVDF as a function of environment, they did examine the effects of cosmic
variance and found that there were no significant differences in the slope of the CVDF between different quadrants of the sky.   The standard way to explain the differences 
between the predictions of simulations and theory and the observed CVDF is the inclusion of the proper baryon physics.  The clear difference between the CVDF
in clusters and the CVDF of the field and groups provide an additional constraint on the possible explanations between observations and theory.  Explanations of the difference
between theory and observations include:  dwarf galaxies inhabit only the most massive haloes today \citep{stoehr02} or only the most massive halos when they were accreted
by a larger halo \citep{kravtsov04}, or only those that collapsed before re-ionization \citep[][cf. Fenner \etal\ 2006]{bullock00}.  If, however, the discrepancy is due to warm instead 
of cold dark matter, this should be independent of environment.  In their study of the ``missing satellite" problem, \citet{simon07} found the best match with their data for Local Group 
dwarfs if they only considered those dwarf galaxy halos that collapsed before reionization.  If this explanation holds for our sample, then low mass halos in clusters must have 
collapsed before those in groups, as would be expected for higher density regions with shorter dynamical times.  

\section{Conclusions}
\label{sec:conclusions}

We have conducted an \HI\ survey of the entire area of six loose groups that are analogous to the Local Group using the Parkes multibeam receiver and the ATCA.   Our survey 
has two goals: (i) to compare the \HI\ properties of loose groups to the Local Group and other groups, and (ii) to examine how the HIMF and CVDF of loose groups compare to 
those in the Local Group, other environments, and simulations.  

We found the following:

\begin{itemize}
\item Our survey found 61 \HI-rich galaxies in the six groups down to \mhi\ of 9$\times$10$^6$\msun, roughly doubling the number of group galaxies as determined by 
\citet{garcia93}.  All of the \HI-detected objects have properties consistent with gas-rich spiral, irregular, or dwarf irregular galaxies.  The derived masses of the
groups surveyed, including the new detections are all within an order of magnitude of the Local Group.  All the groups have similar radial extent and mean separation of the
large galaxies.  

\item The HIMF of these loose groups has a flat low \mhi\ slope that agrees very well with the HIMF of the Local Group.  Both are flatter than the HIMF of field galaxies 
from HIPASS \citep{zwaan05} and are consistent with the idea that the HIMF flattens as the local galaxy density decreases.

\item The CVDF of loose groups agrees very well with that of the Local Group \HI-detected galaxies, although it is lower than the CVDF for all Local Group galaxies.  
The loose group CVDF has the same low-V$_{rot}$ slope as was found in an optical study of field galaxies by \citet{gonzalez00} and for HIPASS galaxies by \citet{zwaan10}. 
The loose group CVDF low-V$_{rot}$ slope is significantly flatter than that of cluster galaxies \citep{desai04} or predicted by dark matter only simulations of Milky Way-sized
halos \citep{diemand08}.  Only in dense clusters are their significant differences in the CVDF from the field or groups.

\item Overall, our survey shows that the Local Group is not atypical in terms of the \HI\ properties of its galaxies nor the properties of the dark matter halos hosting \HI-rich
galaxies.  
\end{itemize}

While our survey has provided a measurement of the HIMF and CVDF in the loose group environment, it is based on a relatively small number of galaxies, only 61.  The 
restricted range of group density and morphology probed by previous studies and the small number of group members have been the main limitations of past studies, 
including our own.  Fortunately, currently ongoing and planned \HI\ surveys will help improve this situation in a variety of ways.  ALFALFA \citep{giovanelli05} is detecting
galaxies out to larger distances and down to lower \mhi\ than HIPASS, but only 40\% of the survey has been used for the published HIMF and CVDF \citep{martin10,papastergis11}.  
ALFALFA allows the study of relatively massive galaxy properties over a range of environments, however AGES \citep{auld06}, is studying a wide 
range of environments from isolated galaxies through galaxy groups to galaxy clusters down to lower \mhi.  In the future, planned \HI\ surveys with SKA pathfinder 
instruments, such as WALLABY with ASKAP\footnote{\url http://www.atnf.csiro.au/research/WALLABY} or with APERTIF on WSRT \citep{oosterloo10}, will further improve our 
understanding of how \HI\ properties of galaxies vary with environment.   

\acknowledgements

The authors wish to thank the staff at Parkes and the ATCA for their 
assistance with observing.  We thank Warwick Wilson for his excellent 
work in making the 16 MHz filters for these observations.  We also thank the
anonymous referee for his/her prompt review and helpful comments which
improved this paper.  This research 
was performed in part while D.J.P. held a National Research Council Research 
Associateship Award at the Naval Research Laboratory.  D.J.P. also 
acknowledges generous support from NSF MPS International Distinguished 
Research Fellowship grant AST 0104439 and partial support from an ATNF 
Bolton Fellowship.  B.K.G. acknowledges the generous financial support provided by
Saint Mary University's Visitor Program.  The authors also wish to thank Bill Saxton for his assistance in 
making the group figures.  We acknowledge the usage of the HyperLeda database ({\url http://leda.univ-lyon1.fr}).
This research has made use of the NASA/IPAC Extragalactic Database (NED) which is operated by the Jet Propulsion Laboratory, 
California Institute of Technology, under contract with the National Aeronautics and Space Administration.

\clearpage

\begin{deluxetable}{lcccccc}
\tablecolumns{7}
\tablewidth{0pc}
\tablecaption{Group Galaxy \HI\ Detections\label{tab:grpdet}}
\tablehead{\colhead{Group} & \colhead{Galaxy\tablenotemark{a}} & 
\colhead{$\alpha$ (2000)\tablenotemark{b}} & 
\colhead{$\delta$ (2000)\tablenotemark{b}} &
\colhead{V$_\odot$\tablenotemark{c}}  &
\colhead{W$_{20}$\tablenotemark{c}} &
\colhead{S$_{int}$\tablenotemark{c}}\\
\colhead{} & \colhead{} & \colhead{} & \colhead{} &\colhead{\kms} & \colhead{\kms} & \colhead{Jy~\kms}}
\startdata
LGG 93        & NGC 1311          & 03 20 07.4 & -52 11 17 &  573$\pm$ 2     & 104$\pm$ 4     &  14.0$\pm$0.1     \\
              & IC 1959           & 03 33 11.8 & -50 24 31 &  639$\pm$ 1     & 149$\pm$ 2     &  28.2$\pm$0.1     \\
              & ESO 200-G45       & 03 35 01.2 & -51 27 09 & 1026$\pm$ 4     &  52$\pm$ 8     &   3.65$\pm$0.08     \\
              & IC 1914           & 03 19 25.2 & -49 36 11 & 1029$\pm$ 1     & 215$\pm$ 2     &  44.3$\pm$0.1 \\
              & LSBG F200-023     & 03 16 28.4 & -49 24 02 & 1045$\pm$ 5     &  88$\pm$10     &   2.2$\pm$0.2     \\
              & IC 1954           & 03 31 32.1 & -51 54 17 & 1062$\pm$ 2     & 231$\pm$ 4     &  20.2$\pm$0.2     \\
              & IC 1896           & 03 07 52.6 & -54 13 01 & 1076$\pm$10     & 118$\pm$20     &   2.7$\pm$0.2     \\
              & IC 1933           & 03 25 39.5 & -52 47 04 & 1060$\pm$ 4     & 208$\pm$ 8     &  24.6$\pm$0.1     \\
              & NGC 1249          & 03 10 04.6 & -53 20 01 & 1073$\pm$ 1     & 238$\pm$ 2     &  99.1$\pm$0.2     \\
              & AM 0311-492       & 03 12 42.8 & -49 10 58 & 1308$\pm$ 5     &  64$\pm$10     &   1.6$\pm$0.1     \\
LGG 106      & APMUKS B0403-4939 & 04 04 38.0 & -49 30 52 &  854$\pm$ 2     &  43$\pm$ 4     &   0.90$\pm$0.07     \\
              & ESO 249-G36       & 03 59 15.6 & -45 52 14 &  898$\pm$ 1     &  80$\pm$ 2     &  15.41$\pm$0.09     \\
              & IC 2000           & 03 49 08.0 & -48 51 26 &  980$\pm$ 1     & 280$\pm$ 2     &  31.9$\pm$0.1     \\
              & IC 2004           & 03 51 43.8 & -49 25 12 & 1003$\pm$ 2     & 126$\pm$ 4     &   1.3$\pm$0.1     \\
              & AM 0358-465       & 03 59 56.6 & -46 46 58 & 1006$\pm$ 2     &  84$\pm$ 4     &   3.71$\pm$0.09     \\
              & ESO 249-G35       & 03 58 56.3 & -45 51 35 & 1030$\pm$ 3     & 128$\pm$ 6     &   5.3$\pm$0.1     \\
              & NGC 1433          & 03 42 00.4 & -47 13 27 & 1076$\pm$ 1     & 184$\pm$ 2     &  31.1$\pm$0.1     \\
              & 6dF J0351-4635    & 03 51 33.2 & -46 35 49 & 1029$\pm$ 1     &  59$\pm$ 2     &   0.86$\pm$0.09     \\
              & ESO 201-G14       & 04 00 27.5 & -49 01 39 & 1052$\pm$ 2     & 167$\pm$ 4     &   8.2$\pm$0.1     \\
              & NGC 1493          & 03 57 27.9 & -46 12 20 & 1053$\pm$ 1     & 119$\pm$ 2     &  41.2$\pm$0.1     \\
              & NGC 1494          & 03 57 43.7 & -48 54 22 & 1131$\pm$ 1     & 183$\pm$ 2     &  28.9$\pm$0.1     \\
              & NGC 1483          & 03 52 47.3 & -47 28 38 & 1149$\pm$ 1     & 155$\pm$ 2     &  19.5$\pm$0.2     \\
              & ESO 201-G23       & 04 10 52.8 & -47 47 10 & 1197$\pm$ 2     &  85$\pm$ 4     &   2.51$\pm$0.09     \\
              & ESO 249-G32       & 03 57 21.6 & -46 22 05 & 1039$\pm$ 2\tablenotemark{d} &  72$\pm$ 7\tablenotemark{d} &   5.3$\pm$0.1\tablenotemark{d} \\
              & APMUKS B0355-4643 & 03 57 08.2 & -46 35 00 & 1169$\pm$ 2\tablenotemark{d} &  91$\pm$13\tablenotemark{d} &   2.6$\pm$0.1\tablenotemark{d} \\
              & NGC 1448          & 03 44 31.0 & -44 38 34 & 1162$\pm$ 4     & 403$\pm$ 8     &  20.6$\pm$0.3     \\
              & ESO 250-G5        & 04 04 36.6 & -46 02 12 & 1217$\pm$ 2     &  49$\pm$ 4     &   0.25$\pm$0.10     \\
LGG 180       & ESO 373-G7        & 09 32 45.6 & -33 14 40 &  929$\pm$ 2     & 230$\pm$ 4     &  79.8$\pm$0.2     \\
              & ESO 373-G20       & 09 43 36.1 & -32 44 35 &  911$\pm$ 2     &  72$\pm$ 4     &   9.4$\pm$0.1     \\
              & UGCA 168          & 09 33 23.0 & -33 02 03 &  926$\pm$ 2     & 229$\pm$ 4     &  61.9$\pm$0.1     \\
              & ESO 434-G41       & 09 47 43.5 & -31 30 13 &  988$\pm$ 2     & 110$\pm$ 4     &  18.4$\pm$0.1     \\
              & UGCA 182          & 09 45 27.9 & -30 20 34 &  998$\pm$ 1     & 143$\pm$ 2     &  22.7$\pm$0.1     \\
              & ESO 373-G6        & 09 31 51.5 & -34 08 13 & 1048$\pm$ 8     &  93$\pm$16     &   3.7$\pm$0.1     \\
              & ESO 434-G19       & 09 40 44.2 & -32 13 45 & 1033$\pm$10     & 129$\pm$ 8     &   5.1$\pm$0.1     \\
              & ESO 434-G17       & 09 37 57.4 & -32 17 20 & 1132$\pm$ 4     &  98$\pm$ 8     &   4.88$\pm$0.09     \\
              & NGC 2997          & 09 45 43.8 & -31 11 59 & 1089$\pm$ 1     & 270$\pm$ 2     & 191.2$\pm$0.2     \\
              & UGCA 177          & 09 44 04.1 & -32 10 07 & 1212$\pm$ 2     &  97$\pm$ 4     &   7.0$\pm$0.2     \\
              & IC 2507           & 09 44 33.9 & -31 47 19 & 1248$\pm$ 7\tablenotemark{d} & 153$\pm$26\tablenotemark{d} &  20.8$\pm$0.1\tablenotemark{d} \\
              & UGCA 180          & 09 44 46.8 & -31 49 13 & 1250$\pm$ 3\tablenotemark{d} & 149$\pm$ 3\tablenotemark{d} &  33.2$\pm$0.1\tablenotemark{d} \\
LGG 293      & APMUKS B1237-0648 & 12 39 44.7 & -07 05 23 &  928$\pm$ 2     &  87$\pm$ 4     &   2.67$\pm$0.09     \\
              & UGCA 289          & 12 35 37.1 & -07 52 22 &  988$\pm$ 2     & 174$\pm$ 4     &  26.9$\pm$0.1     \\
              & NGC 4487          & 12 31 05.3 & -08 03 07 & 1036$\pm$ 2     & 213$\pm$ 4     &  36.6$\pm$0.1     \\
              & NGC 4504          & 12 32 18.9 & -07 33 55 &  999$\pm$ 2     & 243$\pm$ 4     &  93.9$\pm$0.1     \\
              & NGC 4597          & 12 40 11.7 & -05 48 17 & 1036$\pm$ 2     & 183$\pm$ 4     &  61.4$\pm$0.1     \\
              & [KKS2000] 30      & 12 37 36.3 & -08 52 01 & 1101$\pm$ 1     &  49$\pm$ 2     &   1.7$\pm$0.1     \\
              & LCRSB1223-0616    & 12 25 38.7 & -06 33 30 & 1244$\pm$ 2\tablenotemark{d} &  54$\pm$ 3\tablenotemark{d} &   3.0$\pm$0.1\tablenotemark{e} \\
              & LCRSB1223-0612    & 12 25 50.5 & -06 29 24 & 1211$\pm$ 2\tablenotemark{d} &  50$\pm$ 7\tablenotemark{d} &   1.3$\pm$0.1\tablenotemark{e} \\
              & UGCA 286          & 12 33 37.7 & -04 53 12 & 1290$\pm$ 1     & 148$\pm$ 2     &  20.1$\pm$0.2     \\
              & UGCA 295          & 12 44 55.0 & -09 07 27 & 1380$\pm$ 4     & 115$\pm$ 8     &   8.3$\pm$0.3     \\
              & APMUKS B1224-0437 & 12 27 29.2 & -04 53 45 & 1406$\pm$ 2     &  60$\pm$ 4     &   2.2$\pm$0.2     \\
              & DDO 142           & 12 44 04.1 & -05 40 49 & 1430$\pm$ 1     & 136$\pm$ 2     &  38.0$\pm$0.2     \\
              & DDO 146           & 12 45 41.1 & -06 04 26 & 1476$\pm$ 1     & 160$\pm$ 2     &  16.3$\pm$0.2     \\
LGG 478      & APMUKS B2332-3729 & 23 35 05.2 & -37 13 19 &  615$\pm$ 2     &  49$\pm$ 4     &   1.14$\pm$0.07     \\
              & ESO 348-G9        & 23 49 24.7 & -37 46 22 &  649$\pm$ 1     & 101$\pm$ 2     &  13.5$\pm$0.1     \\
              & NGC 7713          & 23 36 14.4 & -37 56 07 &  696$\pm$ 2     & 210$\pm$ 4     &  62.4$\pm$0.2     \\
              & IC 5332           & 23 34 27.4 & -36 06 23 &  701$\pm$ 1     & 117$\pm$ 2     & 168.1$\pm$0.2     \\
              & ESO 347-G17       & 23 26 56.3 & -37 20 35 &  694$\pm$ 1     &  88$\pm$ 2     &   9.32$\pm$0.08     \\
HIPASS Group & NGC 5068          & 13 18 53.8 & -21 02 41 &  670$\pm$ 1     & 110$\pm$ 2     & 133.5$\pm$0.2     \\
              & UGCA 320          & 13 00 36.8 & -17 09 06 &  742$\pm$ 2     & 126$\pm$ 4     & 107.3$\pm$0.3     \\
              & SGC 1257-1909     & 12 59 56.0 & -19 24 29 &  828$\pm$ 1     &  52$\pm$ 2     &   4.5$\pm$0.2     \\
              & MCG-3-34-2        & 13 07 56.6 & -16 41 20 &  958$\pm$ 4     &  53$\pm$ 8     &   1.1$\pm$0.1     \\
\enddata
\tablenotetext{a}{These are the names of the optical counterparts to the \HI\ detection based on a search of NED.}
\tablenotetext{b}{Positions are from the ATCA data and have uncertainties of about 10$\arcsec$.}
\tablenotetext{c}{Data are from Parkes \HI\ spectra, except where noted otherwise.}
\tablenotetext{d}{From ATCA spectrum}
\tablenotetext{e}{Parkes \HI\ flux scaled by ratio of ATCA fluxes}
\end{deluxetable}

\clearpage

\begin{deluxetable}{lcccccc}
\tablecolumns{7}
\tablewidth{0pc}
\tablecaption{Background Galaxy \HI\ Detections\label{tab:bgdet}}
\tablehead{\colhead{Foreground Group} & \colhead{Galaxy\tablenotemark{a}} & 
\colhead{$\alpha$ (2000)\tablenotemark{b}} & 
\colhead{$\delta$ (2000)\tablenotemark{b}} &
\colhead{V$_\odot$\tablenotemark{c}}  &
\colhead{W$_{20}$\tablenotemark{c}} &
\colhead{S$_{int}$\tablenotemark{c}} }
\startdata

LGG 106      & ESO 201-G2        & 03 48 42.7 & -48 25 08 & 1476$\pm$2  & 45$\pm$4 & 1.9$\pm$0.1 \\
             & IC 1986           & 03 40 34.7 & -45 21 19 & 1551$\pm$2 & 112 $\pm$4 & 11.4$\pm$0.2 \\
             & LSBG F249-040     & 03 49 34.9 & -46 34 15 & 1581$\pm$4 & 51$\pm$8 & 0.63$\pm$0.15 \\
             & IC 2009           & 03 53 34.6 & -48 59 31 & 1574$\pm$ 2 & 106$\pm$4 & 6.87$\pm$0.09 \\             
LGG 293      & APMUKS B1237-0724 & 12 40 16.6 & -07 41 05 & 2219$\pm$ 4 &  131$\pm$ 8  & 2.9$\pm$0.1 \\
             & APMUKS B1236-0417 & 12 39 01.2 & -04 33 35 & 2413$\pm$2  &  81$\pm$4 &  5.6$\pm$0.2 \\
             & NGC 4602          & 12 40 38.1 & -05 07 49 &   2539$\pm$4  & 432$\pm$8  & 35.1$\pm$0.2 \\
             & FGC 1496          & 12 44 18.9 & -05 32 12 & 2678$\pm$4 &  209$\pm$8 &  8.2$\pm$ 0.2 \\
             & NGC 4626          & 12 42 27.7 & -06 58 10 & 2800$\pm$8\tablenotemark{d} &  389$\pm$16\tablenotemark{d} & $>$13.2$\pm$0.2\tablenotemark{d} \\
             & HIPASS J1244-08     & 12 45 12.5 & -08 21 09 & 2886 $\pm$2  &  89$\pm$4  & 6.1$\pm$0.1 \\
             & NGC 4433          & 12 27 38.9 & -08 16 39 & $\sim$2977\tablenotemark{d} & $\gtrsim$200\tablenotemark{d}   & $>$8.3$\pm$0.1\tablenotemark{d} \\
LGG 478      & ESO 347-G29       & 23 36 28.2 & -38 47 15 & 1569$\pm$1  & 221$\pm$2 & 30.8$\pm$0.1 \\
             & NGC 7764          & 23 50 54.0 & -40 43 59 &  1668$\pm$8 &  213$\pm$16 &  9.4$\pm$0.7 \\
             & APMUKS B2341-3703 & 23 44 13.6 & -36 46 26 & 1854$\pm$4 &   57$\pm$ 8 &  1.35$\pm$0.09 \\
             & APMUKS B2347-3649 & 23 50 33.7 & -36 33 11 & 2168$\pm$2 &   95$\pm$4 &  0.93$\pm$0.12 \\
             & ESO 408-G12       & 23 37 36.4 & -36 59 04 & 2983$\pm$1 &  201$\pm$2 &  7.3$\pm$0.1 \\
             & ESO 347-G23       & 23 34 31.2 & -39 31 58 & 3028$\pm$4 &  144$\pm$ 8 &  4.5$\pm$0.1 \\
             & NGC 7713A         & 23 37 09.8 & -37 42 56 & 3002$\pm$2 &  152$\pm$ 4 &  7.2$\pm$0.1 \\
HIPASS Group & 2MASX J1314-2203  & 13 14 51.7 & -22 04 30 & 1384$\pm$4 & 173$\pm$ 8 &  6.5$\pm$0.3 \\
             & UGCA 356          & 13 26 36.0 & -22 14 04 & 1418$\pm$4 & 133$\pm$8  & 8.7$\pm$0.2 \\
             & DDO 164           & 13 06 17.9 & -17 30 48 & 1470$\pm$1 &  95$\pm$ 2 & 11.0$\pm$0.2 \\
             & MCG-3-34-67       & 13 24 15.4 & -16 42 16 & 1494$\pm$ 3 & 101$\pm$6 &  4.3$\pm$0.2 \\
             & NGC 5170          & 13 29 48.2 & -17 57 40 & 1502$\pm$1 & 527$\pm$2 & 79.3$\pm$0.6 \\
             & LEDA 083827       & 13 14 30.6 & -16 22 30 & 1487$\pm$2 & 153$\pm$ 4 &  4.9$\pm$0.2 \\
             & ESO 576-G25       & 13 18 29.8 & -20 41 08 & 1560$\pm$ 2 &  79$\pm$4  & 3.2$\pm$0.2 \\
             & ESO 575-G61       & 13 08 14.9 & -20 59 58 & 1642$\pm$3 & 168$\pm$ 6  & 3.3$\pm$0.2 \\
             & NGC 5054          & 13 16 58.7 & -16 38 36 & 1742$\pm$ 2 & 330$\pm$4 & 23.5$\pm$0.2 \\
             & UGCA 348          & 13 19 51.1 & -22 16 38 & 1617$\pm$ 8\tablenotemark{e} & 177$\pm$16\tablenotemark{e}  & 9.2$\pm$0.5\tablenotemark{e} \\
             & NGC 5134          & 13 25 18.6 & -21 08 09 & 1758$\pm$2  & 150$\pm$4  & 9.4$\pm$0.2 \\
             & NGC 5084          & 13 20 15.6 & -21 49 53 &  1715$\pm$4 & 683$\pm$8 & 106.0$\pm$0.4 \\
             & 2MASX J1324-2015  & 13 24 54.4 & -20 17 45 & 1732$\pm$ 2  & 43$\pm$4 &  0.53$\pm$0.15 \\
             & ESO 576-G42       & 13 22 01.9 & -20 13 19 & 1885$\pm$4 & 130$\pm$8  & 4.2$\pm$0.2 \\
             & UGCA 353          & 13 24 41.8 & -19 42 16 & 1964$\pm$4 & 200$\pm$8 & 17.7$\pm$0.2 \\
             & ESO 576-G40       & 13 20 43.6 & -22 03 08 & 1787$\pm$4 & 799$\pm$8 & 50.8$\pm$0.4 \\
             & IC 863            & 13 17 13.2 & -17 15 07 & 2514$\pm$3\tablenotemark{e} & 244$\pm$6\tablenotemark{f}  & 5.9$\pm$0.2\tablenotemark{f} \\        
             & GALEX 2698124594575839357 & 13 17 37.1 & -17 21 41 &2477$\pm$7\tablenotemark{e} & 112$\pm$13\tablenotemark{e}  & 3.2$\pm$0.2\tablenotemark{f}  \\ 
             & SGC 1316-1722     & 13 18 55.5 & -17 38 08 & 2499$\pm$1 & 109$\pm$2 &  4.7$\pm$0.2 \\
             & MCG-3-34-4        & 13 09 43.3 & -16 36 14 & 2569$\pm$4 & 405$\pm$ 8  & 32.6$\pm$0.3 \\
             & ESO 576-G11       & 13 12 54.7\tablenotemark{g}  & -20 01 29\tablenotemark{g}Ê& 2757$\pm$4 & 318$\pm$8 & 17.8$\pm$0.2 \\   
             & ESO 575-G53       & 13 05 05.7 & -22 22 49 & 2644$\pm$8 & 487$\pm$16 & 12.5$\pm$0.5 \\
             & IC 4237           & 13 24 40.1\tablenotemark{g} & -21 10 39\tablenotemark{g} & 2661$\pm$1 & 298$\pm$2  & 11.1$\pm$0.3 \\  
             & LEDA 083801       & 13 13 26.3\tablenotemark{g} & -16 03 30\tablenotemark{g} & 2693$\pm$4  &166$\pm$8 &  7.9$\pm$0.3 \\  
             & SGC 1317-1702     & 13 19 55.0 & -17 18 50 & 2686$\pm$ 2 & 123$\pm$4  & 5.2$\pm$0.1 \\
             & MCG-3-34-14       & 13 12 45.1\tablenotemark{g} & -17 32 21\tablenotemark{g}  & 2763$\pm$4 & 413$\pm$8 & 19.1$\pm$0.3 \\  
             & LEDA 140150       & 13 13 26.3\tablenotemark{g}  & -19 24 21\tablenotemark{g}  & 2780$\pm$8 & 232$\pm$16 &  3.3$\pm$0.3 \\  
             & MCG-3-34-41       & 13 17 06.2 & -16 15 11 & 2636$\pm$2 & 286$\pm$4 &  5.6$\pm$0.3 \\
             & ESO 576-G17       & 13 15 02.3\tablenotemark{g}  & -17 57 25\tablenotemark{g}  & 2771$\pm$1 &  62$\pm$2  & 3.5$\pm$0.1 \\  
             & MCG-3-34-29       & 13 03 11.1\tablenotemark{g} & -17 17 55\tablenotemark{g} & 2966$\pm$1\tablenotemark{d}  & 63$\pm$2\tablenotemark{d}  & $>$4.4$\pm$0.2\tablenotemark{d} \\   
\enddata
\tablenotetext{a}{These are the names of the optical counterparts to the \HI\ detection based on a search of NED.}
\tablenotetext{b}{Positions are from the ATCA data and have uncertainties of about 10$\arcsec$.}
\tablenotetext{c}{Data are from Parkes \HI\ spectra, except where noted otherwise.}
\tablenotetext{d}{\HI\ profile at edge of bandpass, so values are highly uncertain or only lower limits.}
\tablenotetext{e}{From ATCA spectrum}
\tablenotetext{f}{Parkes \HI\ flux scaled by ratio of ATCA fluxes}
\tablenotetext{g}{From Parkes \HI\ data for those galaxies in HICAT and not confirmed by ATCA observations.}
\end{deluxetable}

\clearpage

\begin{deluxetable}{ccccccccc}
\rotate
\tabletypesize{\small}
\tablewidth{0pc}
\tablecolumns{9}
\tablecaption{Group Properties\label{tab:grpprop}}
\tablehead{\colhead{Property} & \colhead{Units} & \colhead{LGG~93} & \colhead{LGG~106} & \colhead{LGG~180} & \colhead{LGG~293} & \colhead{LGG~478} & 
\colhead{HIPASS~Group} & \colhead{Local Group\tablenotemark{a}}}
\startdata
Distance & Mpc & 10.9 & 13.8 & 14.8 & 11.1 & 8.6 & 9.1 & \nodata \\
Number of Members & & 10 & 17 & 12 & 13 & 5 & 4 & 21 \\
V$_\odot$\tablenotemark{b} & \kms & 989 & 1061 & 1064 & 1194 & 671 & 800 & \nodata \\
$\sigma_v$\tablenotemark{c} & \kms & 218 & 101 & 123 & 191 & 38 & 124 & 61$\pm$8\tablenotemark{d} \\
Diameter\tablenotemark{e} & Mpc & 1.0 & 1.4 & 1.2 & 1.2 & 0.68 & 0.92 & 3.8\tablenotemark{f} \\
Mean Galaxy-Galaxy Separation & Mpc & 0.57 & 0.58 & 0.49 & 0.56 & 0.35 & 0.60 & 1.3 \\
$\sigma_r$\tablenotemark{g} & Mpc & 0.48 & 0.48 & 0.46 & 0.45 & 0.27 & 0.46 & 0.53 \\
M$_{vir}$\tablenotemark{h} & 10$^{12}$\msun & 36$\pm$11 & 4.6$\pm$1.1 & 7.0$\pm$2.0 & 11$\pm$3 & 0.13$\pm$0.06 & 0.51$\pm$0.25 & 2.3$\pm$0.6\tablenotemark{d} \\
R$_{0,vir}$\tablenotemark{i} & Mpc & 2.9 & 1.5 & 1.7 & 2.0 & 0.45 & 0.70 & 1.15$\pm$0.15\tablenotemark{d} \\
M$_{pm}$\tablenotemark{j} & 10$^{12}$\msun & 38$\pm$12 & 12$\pm$3 & 10$\pm$3 & 34$\pm$9 & 0.41$\pm$0.18 & 14$\pm$7 & \nodata \\
R$_{0,pm}$\tablenotemark{k} & Mpc & 3.0 & 2.0 & 1.9 & 2.8 & 0.65 & 2.1 & \nodata \\
\enddata
\tablenotetext{a}{only those group members with \mhi$\ge$10$^7$\msun\ are used to calculate number of members and radii of the Local Group.}
\tablenotetext{b}{The mean recession velocity of the group members.}
\tablenotetext{c}{The rms velocity dispersion of the group members.}
\tablenotetext{d}{From \citet{vdb00}.}
\tablenotetext{e}{Twice the projected separation of the most distant group member from the group center.}
\tablenotetext{f}{The radial separation of GR8 from the Local Group barycenter.}
\tablenotetext{g}{The rms dispersion of the projected radial separations of group galaxies.}
\tablenotetext{h}{Calculated using Equation 1 from \citet{heisler85}.}
\tablenotetext{i}{Calculated using Equation 3 from \citet{sandage86} and M$_{vir}$.}
\tablenotetext{j}{Calculated using Equation 2 from \citet{heisler85}.}
\tablenotetext{k}{Calculated using Equation 3 from \citet{sandage86} and M$_{pm}$.}
\end{deluxetable}

\clearpage

\begin{deluxetable}{cccccc}
\tablewidth{0pc}
\tablecolumns{6}
\tablecaption{Interferometer Data for Group Galaxies\label{tab:grpcubes}}
\tablehead{\colhead{Galaxy} &  \colhead{Beam Size} & \colhead{Channel width} &
\multicolumn{2}{c}{Noise} & \colhead{Contour levels\tablenotemark{a}} \\
\colhead{} & \colhead{arcsec} & \colhead{\kms} & \colhead{mJy/beam} & \colhead{10$^{19}$cm$^{-2}$}
& \colhead{10$^{19}$cm$^{-2}$}}
\startdata

NGC 1311           & 76$\times$50 & 6.6 & 5.9 & 1.1 & 5,10,20,50,100  \\
IC 1959      	   & 98$\times$83 & 6.6 & 3.4 & 0.3 & 1,2,5,10,20,50,100 \\
ESO 200-G45	   & 75$\times$51 & 6.6 & 3.5 & 0.7 & 2,5,10,20,50 \\
IC 1914            & 78$\times$54 & 6.6 & 4.0 & 0.7 & 2,5,10,20,50,100 \\
LSBG F200-023	   & 78$\times$54 & 6.6 & 4.0 & 0.7 & 2,5,10,20,50,100 \\
IC 1954            & 75$\times$51 & 6.6 & 3.7 & 0.7 & 2,5,10,20,50,100 \\
IC 1896            & 73$\times$51 & 6.6 & 3.7 & 0.7 & 2,5,10,20 \\
IC 1933            & 75$\times$51 & 6.6 & 3.8 & 0.7 & 2,5,10,20,50,100 \\
NGC 1249      	   & 76$\times$50 & 6.6 & 5.4 & 1.0 & 5,10,20,50,100 \\
AM 0311-492 	   & 78$\times$54 & 6.6 & 4.0 & 0.7 & 2,5,10,20,50,100 \\
APMUKS B0403-4939  & 73$\times$54 & 3.3 & 4.9 & 0.5 & 2,3,4,5 \\
ESO 249-G36	   & 86$\times$57 & 3.3 & 3.7 & 0.3 & 1,2,5,10,20,50,100 \\
ESO 249-G35	   & 86$\times$57 & 3.3 & 3.7 & 0.3 & 1,2,5,10,20,50,100 \\
IC 2000            & 75$\times$63 & 3.3 & 4.0 & 0.3 & 1,2,5,10,20,50,100 \\
IC 2004            &147$\times$98 & 3.3 & 5.1 & 0.1 & 0.5,1 \\
AM 0358-465 	   & 73$\times$57 & 3.3 & 6.0 & 0.5 & 2,6,10,20 \\
NGC 1433      	   & 78$\times$66 & 6.6 & 1.2 & 0.2 & 0.5,1,2,5,10,20 \\
6dF J0351-4635	   &161$\times$130& 3.3 & 5.3 & 0.1 & 0.5,1 \\
ESO 201-G14        & 74$\times$54 & 3.3 & 7.1 & 0.6 & 2,5,10,20 \\
NGC 1493           & 82$\times$51 & 3.3 & 5.8 & 0.5 & 2,5,10,20,50,100 \\
NGC 1494           & 74$\times$53 & 3.3 & 6.5 & 0.6 & 2,5,10,20,50,100 \\
NGC 1483           & 73$\times$56 & 3.3 & 5.8 & 0.5 & 2,5,10,20,50,100 \\
ESO 201-G23   	   & 77$\times$52 & 3.3 & 4.8 & 0.4 & 1,2,5,10 \\
ESO 249-G32   	   & 50$\times$21 & 3.3 & 4.8 & 1.7 & 5,10,20,50,100,200 \\
APMUKS B0355-4643  & 50$\times$21 & 3.3 & 4.8 & 1.7 & 5,10,20,50,100,200 \\
NGC 1448           &157$\times$103& 6.6 & 1.7 & 0.1 & 0.5,1,2,5,10,20,50,100,200 \\
ESO 250-G5     	   &149$\times$98 & 3.3 & 4.9 & 0.1 & 0.5,1,2,5,10 \\
ESO 373-G7     	   & 82$\times$55 & 3.3 & 8.5 & 0.7 & 2,5,10,20,50,100,200  \\
ESO 373-G20   	   & 89$\times$51 & 3.3 & 4.1 & 0.3 & 1,2,5,10,20 \\
UGCA 168           & 82$\times$55 & 3.3 & 8.5 & 0.7 & 2,5,10,20,50,100,200 \\
ESO 434-G41   	   & 82$\times$55 & 3.3 & 10.0& 0.8 & 2,5,10,20,50,100 \\
UGCA 182           & 93$\times$51 & 3.3 & 4.1 & 0.3 & 1,2,5,10,20,50,100 \\
ESO 373-G6         &102$\times$81 & 3.3 & 3.4 & 0.2 & 1,2,5,10,20 \\
ESO 434-G19    	   & 99$\times$49 & 3.3 & 4.7 & 0.5 & 2,5,10,20 \\
ESO 434-G17    	   & 87$\times$55 & 3.3 & 7.0 & 0.5 & 2,5,10,20 \\
NGC 2997\tablenotemark{b} & 36$\times$29 & 6.6 & 0.5 & 0.35 & 1,2,5,10,20,50,100 \\
UGCA 177           & 82$\times$55 & 3.3 & 9.6 & 0.8 & 2,5,10,20,50,100 \\
IC 2507            & 82$\times$55 & 3.3 & 8.0 & 0.6 & 2,5,10,20,50,100 \\
UGCA 180     	   & 82$\times$55 & 3.3 & 8.0 & 0.6 & 2,5,10,20,50,100 \\
APMUKS B1237-0648  &155$\times$124 & 3.3 & 4.1 & 0.08 & 0.5,1,2,5 \\
UGCA 289           &464$\times$349 & 6.6 & 5.7 & 0.03 & 0.1,0.2,0.5,1,2,5,10 \\
NGC 4487           &160$\times$125 & 3.3 & 4.4 & 0.08 & 0.5,1,2,5,10,20,50 \\
NGC 4504           &162$\times$121 & 3.3 & 3.7 & 0.07 & 0.5,1,2,5,10,20,50 \\
NGC 4597           &158$\times$125 & 3.3 & 4.2 & 0.08 & 0.5,1,2,5,10,20,50,100 \\
\,[KKS2000] 30       &157$\times$121 & 3.3 & 4.2 & 0.08 & 0.5,1,2,5 \\
LCRSB1223-0616	   &154$\times$125 & 3.3 & 4.5 & 0.08 & 0.5,1,2,5,10 \\
LCRSB1223-0612	   &154$\times$125 & 3.3 & 4.5 & 0.08 & 0.5,1,2,5,10 \\
UGCA 286           &157$\times$125 & 3.3 & 4.4 & 0.08 & 0.5,1,2,5,10,20,50 \\
UGCA 295           &155$\times$122 & 3.3 & 4.3 & 0.08 & 0.5,1,2,5,10,20 \\
APMUKS B1224-0437  &156$\times$125 & 3.3 & 4.2 & 0.08 & 0.5,1,2,5 \\
DDO 142            &155$\times$126 & 3.3 & 4.6 & 0.09 & 0.5,1,2,5,10,20 \\
DDO 146            & 73$\times$53  & 5.2 & 1.5 & 0.2 &  0.5,1,2,5,10,20,50 \\
APMUKS B2332-3729  &123$\times$47 & 3.3 & 4.1 & 0.3 & 1,2,5 \\
ESO 348-G9         & 90$\times$60 & 3.3 & 3.3 & 0.2 & 0.5,1,2,5,10,20 \\
NGC 7713           &111$\times$48 & 3.3 & 5.0 & 0.3 & 1,2,5,10,20,50,100,200 \\
IC 5332            & 97$\times$61 & 3.3 & 3.4 & 0.2 & 0.5,1,2,5,10,20,50 \\
ESO 347-G17	   &112$\times$50 & 3.3 & 5.6 & 0.4 & 1,2,5,10,20,50 \\
NGC 5068      	   &680$\times$72 &13.2 & 3.6 & 0.1 & 0.5,1,2,5,10,20,50 \\
UGCA 320      	   & 25$\times$18 & 2.6 & 1.1 & 0.7 & 2,5,10,20,50,100,200,500 \\
SGC 1257-1909	   &177$\times$43 & 3.3 & 6.9 & 0.3 & 1,2,5,10,20 \\
MCG-3-34-2         &208$\times$42 & 3.3 & 5.3 & 0.2 & 0.5,1,2,5,10 \\
\enddata
\tablenotetext{a}{Corresponding to Figures~\ref{fig:lgg93_opt}-\ref{fig:hgrp3_opt}.}
\tablenotetext{b}{Data taken from \citet{hess09}.}
\end{deluxetable}

\clearpage

\begin{deluxetable}{cccccc}
\tablewidth{0pc}
\tablecolumns{6}
\tablecaption{Interferometer Data for Background Galaxies\label{tab:bgcubes}}
\tablehead{\colhead{Galaxy} &  \colhead{Beam Size} & \colhead{Channel width} &
\multicolumn{2}{c}{Noise} & \colhead{Contour levels\tablenotemark{a}} \\
\colhead{} & \colhead{arcsec} & \colhead{\kms} & \colhead{mJy/beam} & \colhead{10$^{19}$cm$^{-2}$}
& \colhead{10$^{19}$cm$^{-2}$}}
\startdata
ESO 201-G2        &  75$\times$53 & 3.3 & 6.5 & 0.6 & 2,5,10 \\ 
IC 1986           & 102$\times$49 &13.2 & 4.5 & 1.3 & 5,10,20,50 \\ 
LSBG F249-040     & 178$\times$123& 3.3 & 5.0 & 0.08& 0.5,1,2,5,10 \\ 
IC 2009           &  75$\times$53 & 3.3 & 6.1 & 0.6 & 2,5,10,20 \\ 
APMUKS B1237-0724 & 157$\times$125 & 3.3 & 4.1 & 0.08& 0.5,1,2,5,10 \\ 
APMUKS B1236-0417 & 197$\times$116 & 3.3 & 6.3 & 0.1 & 0.5,1,2,5,10 \\ 
NGC 4602          & 204$\times$116 & 3.3 & 6.6 & 0.1 & 0.5,1,2,5,10,20 \\ 
FGC 1496          & 153$\times$125 & 3.3 & 4.3 & 0.08& 0.5,1,2,5,10,20 \\ 
NGC 4626          & 209$\times$115 & 3.3 & 6.1 & 0.09& 0.5,1,2,5,10,20 \\ 
LGG 293-HI-10     & 212$\times$114 & 3.3 & 6.6 & 0.1 & 0.5,1,2,5,10 \\ 
NGC 4433          & 202$\times$114 & 3.3 & 6.0 & 0.09& 0.5,1,2,5 \\ 
ESO 347-G29       &  63$\times$30 & 6.6 & 1.6 & 0.6 & 2,5,10,20,50,100 \\ 
NGC 7764          &  87$\times$52 & 3.3 & 6.0 & 0.5 & 2,5,10,20,50 \\ 
APMUKS B2341-3703 & 101$\times$55 & 3.3 & 3.3 & 0.2 & 0.5,1,2,5,10 \\ 
APMUKS B2347-3649 & 109$\times$53 & 3.3 & 3.6 & 0.2 & 0.5,1,2,5,10 \\ 
ESO 408-G12       & 123$\times$45 & 3.3 & 5.3 & 0.3 & 1,2,3 \\ 
ESO 347-G23       & 109$\times$49 & 3.3 & 4.4 & 0.3 & 1,2,5,10 \\ 
NGC 7713A         & 124$\times$48 & 3.3 & 5.1 & 0.3 & 1,2,5,10,20 \\ 
2MASX J1314-2203  & 155$\times$43 & 3.3 & 5.6 & 0.3 & 1,2,5,10 \\ 
UGCA 356          & 158$\times$42 & 3.3 & 7.0 & 0.4 & 1,2,5,10,20 \\ 
DDO 164           & 225$\times$39 & 3.3 & 7.0 & 0.3 & 1,2,5,10,20 \\ 
MCG-3-34-67       & 192$\times$46 & 3.3 & 5.8 & 0.2 & 0.5,1,2,5,10,20 \\ 
NGC 5170          & 170$\times$49 & 3.3 & 8.0 & 0.3 & 1,2,5,10,20,50,100 \\ 
LEDA 083827       & 205$\times$44 & 3.3 & 5.3 & 0.2 & 0.5, 1,2,5,10 \\ 
ESO 576-G25       & 152$\times$47 & 3.3 & 7.6 & 0.4 & 1,2,5,10 \\ 
ESO 575-G61       & 123$\times$50 & 6.6 & 2.5 & 0.3 & 1,2,5,10 \\ 
NGC 5054          & 193$\times$46 & 3.3 & 7.6 & 0.3 & 1,2,5,10,20 \\ 
UGCA 348          & 132$\times$50 & 3.3 & 6.3 & 0.3 & 1,2,5,10,20 \\ 
NGC 5134          & 143$\times$48 & 3.3 & 7.6 & 0.4 & 1,2,5,10,20 \\ 
NGC 5084          & 222$\times$72 & 6.6 & 4.4 & 0.2 & 0.5,1,2,5,10,20,50 \\ 
2MASX J1324-2015  & 165$\times$43 & 3.3 & 5.0 & 0.3 & 1,2,5 \\ 
ESO 576-G42       & 173$\times$42 & 3.3 & 6.5 & 0.3 & 1,2,5,10 \\ 
UGCA 353          & 165$\times$45 & 3.3 & 7.2 & 0.4 & 1,2,5,10,20,50 \\ 
ESO 576-G40       & 154$\times$44 & 3.3 & 6.2 & 0.3 & 1,2,5,10,20,50,100 \\ 
IC 863            & 197$\times$43 & 3.3 & 5.0 & 0.2 & 0.5,1,2,5,10 \\ 
GALEX 2698124594575839357 & 197$\times$43 & 3.3 & 5.0 & 0.2 & 0.5,1,2,5,10 \\ 
SGC 1316-1722     & 212$\times$41 & 3.3 & 7.1 & 0.3 & 1,2,5,10 \\ 
MCG-3-34-4        & 296$\times$37 & 3.3 & 7.8 & 0.3 & 1,2,5,10,20 \\ 
ESO 575-G53       & 143$\times$46 & 3.3 & 6.7 & 0.4 & 1,2,5,10,20 \\ 
SGC 1317-1702     & 189$\times$44 & 3.3 & 6.6 & 0.3 & 1,2,5,10,20 \\ 
MCG-3-34-41       & 202$\times$44 & 3.3 & 6.9 & 0.3 & 1,2,5,10,20 \\ 
\enddata
\tablenotetext{a}{Corresponding to Figures~\ref{fig:lgg106_bg_opt}-\ref{fig:hgrp3_bg_opt}.}
\end{deluxetable}

\clearpage

\begin{deluxetable}{ccccccc}
\tablewidth{0pc}
\tablecolumns{7}
\tablecaption{Derived Properties of Group Galaxies\label{tab:grpgalprops}}
\tablehead{\colhead{Galaxy} & \colhead{Distance\tablenotemark{a}} & \colhead{\mhi} & 
\colhead{inclination\tablenotemark{b}} & \colhead{V$_{rot}$\tablenotemark{c}} & 
\colhead{M$_B$\tablenotemark{d}} & \colhead{\mhi/L$_B$} \\
\colhead{} & \colhead{Mpc} & \colhead{10$^8$\msun} & \colhead{$\arcdeg$} & \colhead{\kms} & 
\colhead{mag} & \colhead{\msun/\lsun}}
\startdata
         NGC~1311 & 10.9 &  3.93 & 90 &  45 & -16.9 &  0.45 \\
          IC~1959 & 10.9 &  7.91 & 90 &  65 & -17.0 &  0.78 \\
      ESO~200-G45 & 10.9 &  1.01 & 37 &  36 & -13.9 &  1.71 \\
          IC~1914 & 10.9 & 12.42 & 43 & 141 & -16.8 &  1.55 \\
    LSBG~F200-023 & 10.9 &  0.62 & 45 &  53 & -12.4 &  4.32 \\
          IC~1954 & 10.9 &  5.66 & 69 & 111 & -18.2 &  0.20 \\
          IC~1896 & 10.9 &  0.76 & 76 &  52 & -15.2 &  0.39 \\
          IC~1933 & 10.9 &  6.90 & 58 & 108 & -17.4 &  0.50 \\
         NGC~1249 & 10.9 & 27.79 & 69 & 115 & -18.1 &  1.07 \\
      AM~0311-492 & 10.9 &  0.45 & 52 &  34 & -12.9 &  2.02 \\
APMUKS~B0403-4939 &13.8 & 0.40 & 71 & 17 & -12.5 & 2.70 \\
      ESO~249-G36 & 13.8 &  6.92 & 42 &  51 & -15.4 &  2.97 \\
          IC~2000 & 13.8 & 14.34 & 90 & 128 & -17.7 &  0.74 \\
          IC~2004 & 13.8 &  0.58 & 44 &  77 & -15.7 &  0.19 \\
      AM~0358-465 & 13.8 &  1.66 & 43 &  53 & -15.8 &  0.49 \\
      ESO~249-G35 & 13.8 &  2.38 & 90 &  55 & -14.5 &  2.34 \\
         NGC~1433 & 13.8 & 13.98 & 67 &  88 & -19.9 &  0.09 \\
   6dF~J0351-4635 & 13.8 &  0.36 & 17 &  82 & -14.8 &  0.27 \\
      ESO~201-G14 & 13.8 &  3.69 & 90 &  73 & -16.6 &  0.52 \\
         NGC~1493 & 13.8 & 18.52 & 23 & 132 & -18.9 &  0.31 \\
         NGC~1494 & 13.8 & 12.99 & 69 &  86 & -18.5 &  0.32 \\
         NGC~1483 & 13.8 &  8.76 & 37 & 112 & -17.5 &  0.55 \\
      ESO~201-G23 & 13.8 &  1.12 & 62 &  40 & -14.2 &  1.55 \\
      ESO~249-G32 & 13.8 &  2.38 & 90 &  30 & -14.4 &  2.57 \\
APMUKS~B0355-4643 & 13.8 &  1.17 & 60 &  45 & -13.7 &  2.40 \\
         NGC~1448 & 13.8 &  9.26 & 86 & 190 & -19.4 &  0.11 \\
       ESO~250-G5 & 13.8 &  0.09 & 60 &  21 & -16.6 &  0.01 \\
       ESO~373-G7 & 14.8 & 41.25 & 66 & 113 & -15.0 & 26.00 \\
      ESO~373-G20 & 14.8 &  4.86 & 47 &  42 & -15.3 &  2.48 \\
         UGCA~168 & 14.8 & 32.00 & 79 & 105 & -18.7 &  0.66 \\
      ESO~434-G41 & 14.8 &  9.51 & 90 &  48 & -16.5 &  1.56 \\
         UGCA~182 & 14.8 & 11.73 & 90 &  62 & -16.7 &  1.63 \\
       ESO~373-G6 & 14.8 &  1.91 & 50 &  53 & -15.2 &  0.99 \\
      ESO~434-G19 & 14.8 &  2.64 & 88 &  56 & -16.2 &  0.57 \\
      ESO~434-G17 & 14.8 &  2.48 & 55 &  51 & -15.5 &  1.01 \\
         NGC~2997 & 14.8 & 98.84 & 32 & 233 & -21.3 &  0.19 \\
         UGCA~177 & 14.8 &  3.62 & 15 & 162 & -16.2 &  0.79 \\
          IC~2507 & 14.8 & 10.75 & 73 &  70 & -18.0 &  0.43 \\
         UGCA~180 & 14.8 & 17.16 & 33 & 120 & -17.9 &  0.75 \\
APMUKS~B1237-0648 & 11.1 &  0.76 & 90 &  37 & -13.9 &  1.32 \\
         UGCA~289 & 11.1 &  7.82 & 90 &  76 & -15.4 &  3.33 \\
         NGC~4487 & 11.1 & 10.64 & 46 & 132 & -18.5 &  0.27 \\
         NGC~4504 & 11.1 & 27.30 & 50 & 143 & -18.1 &  0.98 \\
         NGC~4597 & 11.1 & 17.85 & 90 &  80 & -17.4 &  1.22 \\
     \,[KKS2000]~30 & 11.1 &  0.49 & 66 &  20 & -13.1 &  1.89 \\
   LCRSB1223-0616 & 11.1 &  0.87 & 53 &  28 & -12.6 &  5.18 \\
   LCRSB1223-0612 & 11.1 &  0.38 & 58 &  23 & -12.8 &  1.87 \\
         UGCA~286 & 11.1 &  5.84 & 90 &  64 & -15.2 &  3.13 \\
         UGCA~295 & 11.1 &  2.41 & 22 & 133 & -16.0 &  0.63 \\
APMUKS~B1224-0437 & 11.1 &  0.64 & 77 &  25 & -13.5 &  1.57 \\
          DDO~142 & 11.1 & 11.05 & 27 & 130 & -17.4 &  0.81 \\
          DDO~146 & 11.1 &  4.74 & 52 &  88 & -17.3 &  0.35 \\
APMUKS~B2332-3729 &  8.6 &  0.19 & 62 &  22 & -13.1 &  0.68 \\
       ESO~348-G9 &  8.6 &  2.36 & 90 &  43 & -13.0 &  9.27 \\
         NGC~7713 &  8.6 & 10.89 & 66 & 102 & -18.2 &  0.35 \\
          IC~5332 &  8.6 & 29.34 & 18 & 164 & -18.4 &  0.79 \\
      ESO~347-G17 &  8.6 &  1.62 & 90 &  38 & -14.8 &  1.21 \\
         NGC~5068 &  9.1 & 26.09 & 27 & 105 & -19.6 &  0.23 \\
         UGCA~320 &  9.1 & 20.97 & 90 &  54 & -16.6 &  2.96 \\
    SGC~1257-1909 &  9.1 &  0.88 & 67 &  23 & -13.7 &  1.85 \\
       MCG-3-34-2 &  9.1 &  0.21 & 54 &  27 & -15.3 &  0.10 \\
\enddata
\tablenotetext{a}{Taken from Paper~I.}
\tablenotetext{b}{Taken from Hyperleda, where available, or calculated in same fashion using data from NED.}
\tablenotetext{c}{Calculated from W$_{20}$ using the method described in \citet{meyer08}.}
\tablenotetext{d}{Calculated from apparent B magnitude from HyperLeda or b$_J$ magnitude from NED.  Corrected
for external extinction using \citet{schlegel98}, no internal extinction correction applied.}
\end{deluxetable}

\clearpage

\clearpage

\clearpage

\clearpage

\clearpage

\clearpage

\clearpage

\clearpage

\clearpage

\clearpage

\clearpage

\clearpage

\clearpage

\clearpage

\clearpage

\clearpage

\clearpage

\clearpage

\clearpage

\clearpage

\clearpage

\clearpage

\clearpage

\clearpage

\clearpage

\clearpage

\clearpage

\clearpage

\clearpage

\clearpage

\clearpage

\end{document}